\documentclass[11pt,english]{article}
\usepackage[T1]{fontenc}
\usepackage[latin9]{inputenc}
\usepackage[a4paper]{geometry}
\usepackage{color}
\usepackage{mathrsfs}
\usepackage{mathtools}
\usepackage{amsmath}
\usepackage{amsthm}
\usepackage{amssymb}
\usepackage{graphicx}
\usepackage{setspace}
\usepackage[authoryear]{natbib}
\usepackage{hyperref}
\definecolor{BlueInkPen}{RGB}{0,15,85}
\definecolor{RedCrush}{RGB}{205,83,38}
\hypersetup{
  colorlinks = true,
  citecolor=black,
  linkcolor = BlueInkPen, 
  urlcolor = BlueInkPen,
}

\makeatletter
  \theoremstyle{plain}
  \newtheorem{assumption}{\protect\assumptionname}
  \theoremstyle{plain}
  
\theoremstyle{plain}
\newtheorem{thm}{\protect\theoremname}

\@ifundefined{date}{}{\date{}}
\bibpunct{(}{)}{,}{a}{,}{,}

\usepackage{babel}

\usepackage[multiple]{footmisc}

\usepackage{chngcntr}
\usepackage{apptools}
\AtAppendix{\counterwithin{lemma}{section}}

\makeatother

\usepackage{babel}
  \providecommand{\assumptionname}{Assumption}
  \providecommand{\lemmaname}{Lemma}
\providecommand{\theoremname}{Theorem}

\begin{document}

\title{\vspace{20pt}
A mixture autoregressive model based on Student's $t$--distribution%
\thanks{Contact
addresses: 
Mika Meitz, 
Discipline of Economics, University of Helsinki, P. O. Box 17, FI--00014 University of Helsinki, Finland; 
e-mail: mika.meitz@helsinki.fi. 
Daniel Preve, 
Department of Economics and Finance, City University of Hong Kong, 83 Tat Chee Avenue, Kowloon, Hong Kong, China;
e-mail: pdapreve@cityu.edu.hk.
Pentti Saikkonen, 
Department of Mathematics and Statistics, University of Helsinki, P. O. Box 68, FI--00014 University of Helsinki, Finland;
e-mail: pentti.saikkonen@helsinki.fi.%
}\vspace{20pt}
}

\author{Mika Meitz\\\small{University of Helsinki} \and Daniel Preve\\\small{City University of Hong Kong} \and Pentti Saikkonen\\\small{University of Helsinki}\vspace{20pt}
}

\date{May 2018}
\maketitle
\begin{abstract}
\noindent A new mixture autoregressive model based on Student's $t$\textendash distribution
is proposed. A key feature of our model is that the conditional $t$\textendash distributions
of the component models are based on autoregressions that have multivariate
$t$\textendash distributions as their (low-dimensional) stationary
distributions. That autoregressions with such stationary distributions
exist is not immediate. Our formulation implies that the conditional
mean of each component model is a linear function of past observations
and the conditional variance is also time varying. Compared to previous
mixture autoregressive models our model may therefore be useful in
applications where the data exhibits rather strong conditional heteroskedasticity.
Our formulation also has the theoretical advantage that conditions
for stationarity and ergodicity are always met and these properties
are much more straightforward to establish than is common in nonlinear
autoregressive models. An empirical example employing a realized kernel
series based on S\&P 500 high-frequency data shows that the proposed model performs
well in volatility forecasting.

\bigskip{}
\bigskip{}
\bigskip{}



\end{abstract}
\vfill{}
\pagebreak{}

\section{Introduction}

Different types of mixture models are in widespread use in various
fields. Overviews of mixture models can be found, for example, in
the monographs of \citet{mclachlan2000finite} and \citet{fruhwirth2006finite}.
In this paper, we are concerned with mixture autoregressive models
that were introduced by \citet{le1996modeling} and further developed
by \citet{wong2000mixture,wong2001logistic,wong2001mixture} (for
further references, see \citet{kalliovirta2015gaussian}).

In mixture autoregressive models the conditional distribution
of the present observation given the past is a mixture distribution
where the component distributions are obtained from linear autoregressive
models. The specification of a mixture autoregressive model typically requires 
two choices: choosing a conditional distribution for the component
models and choosing a functional form for the mixing weights. In a majority of existing models a Gaussian
distribution is assumed whereas, in addition to constants, several
different time-varying mixing weights (functions of past observations)
have been considered in the literature. 

Instead of a Gaussian distribution, \citet{wong2009student} proposed
using Student's $t$\textendash distribution. A major motivation for
this comes from the heavier tails of the $t$\textendash distribution
which allow the resulting model to better accommodate for the fat
tails encountered in many observed time series, especially in economics
and finance. In the model suggested by \citet{wong2009student}, the
conditional mean and conditional variance of each component model
are the same as in the Gaussian case (a linear function of
past observations and a constant, respectively), and what changes
is the distribution of the independent and identically distributed
error term: instead of a standard normal distribution, a Student's
$t$\textendash distribution is used. This is a natural approach to
formulate the component models and hence also a mixture autoregressive model based on
the $t$\textendash distribution. 

In this paper, we also consider a mixture autoregressive model based on Student's $t$\textendash distribution,
but our specification differs from that used by \citet{wong2009student}.
Our starting point is the characteristic feature of linear Gaussian
autoregressions that stationary distributions (of consecutive observations)
as well as conditional distributions are Gaussian. We imitate
this feature by using a (multivariate) Student's $t$\textendash distribution
and, as a first step, construct a linear autoregression in which
both conditional and (low-dimensional) stationary distributions 
have Student's $t$\textendash distributions.
This leads to a model
where the conditional mean is as in the Gaussian case (a linear function
of past observations) whereas the conditional variance is no longer constant
but depends on a quadratic form of past observations.
These linear models are then used as component models in our new mixture autoregressive model
which we call the StMAR model.

Our StMAR model has some very attractive features. Like the model of \citet{wong2009student},
it can be useful for modelling time series with regime switching, multimodality,
and conditional heteroskedasticity. As the conditional variances of
the component models are time-varying, the StMAR model can potentially
accommodate for stronger forms of conditional heteroskedasticity than
the model of \citet{wong2009student}. Our formulation also has the
theoretical advantage that, for a $p$th order model, the stationary
distribution of $p+1$ consecutive observations is fully known and
is a mixture of particular Student's $t$\textendash distributions. 
Moreover, stationarity and ergodicity
are simple consequences of the definition of the model and do not
require complicated proofs.

Finally, a few notational conventions. All vectors 
are treated as column vectors and
we write $\boldsymbol{x}=(x_{1},\ldots,x_{n})$ for the
vector $\boldsymbol{x}$ where the components $x_{i}$ may be either
scalars or vectors. The notation $\mathbf{X}\sim n_{d}(\boldsymbol{\mu},\mathbf{\Gamma})$
signifies that the random vector $\mathbf{X}$ has a $d$\textendash dimensional
Gaussian distribution with mean $\boldsymbol{\mu}$ and (positive
definite) covariance matrix $\mathbf{\Gamma}$. Similarly, by $\mathbf{X}\sim t_{d}(\boldsymbol{\mu},\mathbf{\Gamma},\nu)$
we mean that 
$\mathbf{X}$ has a $d$\textendash dimensional
Student's $t$\textendash distribution with mean $\boldsymbol{\mu}$,
(positive definite) covariance matrix $\mathbf{\Gamma}$, and degrees
of freedom $\nu$ (assumed to satisfy $\nu>2$); 
the density function and some properties of the multivariate
Student's $t$\textendash distribution employed are given
in an Appendix. 
The notation $\mathbf{1}_{d}$ is used for a $d$\textendash dimensional
vector of ones, $\imath_{d}$ signifies the vector
$(1,0,\ldots,0)$ of dimension $d$, and the identity matrix of dimension
$d$ is denoted by $I_{d}$. The Kronecker product is denoted by
$\otimes$, and $vec(A)$ stacks the columns of matrix $A$ on top
of one another.

\section{Linear Student's $t$ autoregressions}

In this section we briefly consider linear $p$th order autoregressions
that have multivariate Student's $t$\textendash distributions as their
stationary distributions. First, for motivation and to develop notation,
consider a linear Gaussian autoregression $z_{t}$ ($t=1,2,\ldots$)
generated by 
\begin{equation}
z_{t}=\varphi_{0}+\sum_{i=1}^{p}\varphi_{j}z_{t-i}+\sigma e_{t},\label{eq:LinN_Model}
\end{equation}
where the error terms $e_{t}$ are independent and identically distributed
with a standard normal distribution, and the parameters satisfy
$\varphi_{0}\in\mathbb{R}$, $\boldsymbol{\varphi}=(\varphi_{1},\ldots,\varphi_{p})\in\mathbb{S}^{p}$,
and $\sigma>0$, where 
\begin{equation}
\mathbb{S}^{p}=\{(\varphi_{1},\ldots,\varphi_{p})\in\mathbb{R}^{p}:\varphi\left(z\right)=1-\sum_{i=1}^{p}\varphi_{i}z^{i}\neq0\text{ \ for }\left\vert z\right\vert \leq1\}\label{eq:Sp}
\end{equation}
is the
stationarity region of a linear $p$th order
autoregression. Denoting $\boldsymbol{z}_{t}=(z_{t},\ldots,z_{t-p+1})$
and $\boldsymbol{z}_{t}^{+}=(z_{t},\boldsymbol{z}_{t-1})$,
it is well known that the stationary solution $z_{t}$ to (\ref{eq:LinN_Model})
satisfies
\begin{align}
\boldsymbol{z}_{t} & \sim n_{p}(\mu\mathbf{1}_{p},\mathbf{\Gamma}_{p}),\nonumber \\
\boldsymbol{z}_{t}^{+} & \sim n_{p+1}(\mu\mathbf{1}_{p+1},\mathbf{\Gamma}_{p+1}),\label{eq:LinN_Distr}\\
z_{t}\mid\boldsymbol{z}_{t-1} & \sim n_{1}(\varphi_{0}+\boldsymbol{\varphi}'\boldsymbol{z}_{t-1},\sigma^{2})=n_{1}(\mu+\boldsymbol{\gamma}_{p}'\mathbf{\Gamma}_{p}^{-1}(\boldsymbol{z}_{t-1}-\mu\mathbf{1}_{p}),\sigma^{2}),\nonumber 
\end{align}
where the last relation defines the conditional distribution of $z_{t}$
given $\boldsymbol{z}_{t-1}$ and the quantities $\mathbf{\Gamma}_{p}$,
$\gamma_{0}$, $\boldsymbol{\gamma}_{p}$, $\mu$, and $\mathbf{\Gamma}_{p+1}$
are defined via 
\begin{align}
&vec(\mathbf{\Gamma}_{p})  =(I_{p^{2}}-(\Phi\otimes\Phi))^{-1}\,\imath_{p^{2}}\,\sigma^{2},\quad
\Phi=\begin{bmatrix}\varphi_{1}\cdots\varphi_{p-1} & \varphi_{p}\\ I_{p-1} & \mathbf{0}_{p-1} \end{bmatrix},\nonumber \\
&\gamma_{0}  =\sigma^{2}+\boldsymbol{\varphi}'\mathbf{\Gamma}_{p}\boldsymbol{\varphi},\quad
\boldsymbol{\gamma}_{p}=\mathbf{\Gamma}_{p}\boldsymbol{\varphi},\quad
\mu=\varphi_{0}/(1-\varphi_{1}-\cdots-\varphi_{p}),\quad
\mathbf{\Gamma}_{p+1}  = 
\begin{bmatrix}\mathbf{\gamma}_{0} & \boldsymbol{\gamma}_{p}'\\ \boldsymbol{\gamma}_{p} & \mathbf{\Gamma}_{p} \end{bmatrix}.
\label{eq:NotationGamma}
\end{align}
Two essential properties of linear Gaussian autoregressions are that
they have the distributional features in (\ref{eq:LinN_Distr}) and
the representation in (\ref{eq:LinN_Model}). 

It is not immediately obvious that linear autoregressions based on Student's $t$\textendash distribution with similar properties exist
(such models have, however, appeared at least in \citet{spanos1994modeling} and \citet{heracleous2006student}).
Suppose that for a random vector 
in $\mathbb{R}^{p+1}$ it holds that $(z,\boldsymbol{z})\sim t_{p+1}(\mu\mathbf{1}_{p+1},\mathbf{\Gamma}_{p+1},\nu)$
where $\nu>2$ (and other notation is as above in (\ref{eq:NotationGamma})).
Then (for details, see the Appendix) the conditional distribution of
$z$ given $\boldsymbol{z}$ is $z\mid\boldsymbol{z}\sim t_{1}(\mu(\boldsymbol{z}),\sigma^{2}(\boldsymbol{z}),\nu+p)$,
where
\begin{equation}
\mu(\boldsymbol{z}) = \varphi_{0}+\boldsymbol{\varphi}'\boldsymbol{z},\qquad
\sigma^{2}(\boldsymbol{z}) = \frac{\nu-2+(\boldsymbol{z}-\mu\mathbf{1}_{p})^{\prime}\mathbf{\Gamma}_{p}^{-1}(\boldsymbol{z}-\mu\mathbf{1}_{p})}{\nu-2+p}\sigma^{2}.\label{eq:z_cond_mean_var}
\end{equation}
We now state the following theorem (proofs of all theorems are in the Supplementary Material). 
\begin{thm}
Suppose $\varphi_{0}\in\mathbb{R}$, $\boldsymbol{\varphi}=(\varphi_{1},\ldots,\varphi_{p})\in\mathbb{S}^{p}$,
$\sigma>0$, and $\nu>2$. Then there exists a process $\boldsymbol{z}_{t}=(z_{t},\ldots,z_{t-p+1})$
($t=0,1,2,\ldots$) with the following properties. 

(i) The process $\boldsymbol{z}_{t}$ ($t=1,2,\ldots$) is a
Markov chain on $\mathbb{R}^{p}$ with a stationary distribution characterized
by the density function $t_{p}(\mu\mathbf{1}_{p},\mathbf{\Gamma}_{p},\nu)$.
When $\boldsymbol{z}_{0}\sim t_{p}(\mu\mathbf{1}_{p},\mathbf{\Gamma}_{p},\nu)$,
we have, for $t=1,2,\ldots$, that $\boldsymbol{z}_{t}^{+} \sim t_{p+1}(\mu\mathbf{1}_{p+1},\mathbf{\Gamma}_{p+1},\nu)$
and the conditional distribution of $z_{t}$ given $\boldsymbol{z}_{t-1}$
is 
\begin{equation}
z_{t}\mid\boldsymbol{z}_{t-1}\sim t_{1}(\mu(\boldsymbol{z}_{t-1}),\sigma^{2}(\boldsymbol{z}_{t-1}),\nu+p).\label{eq:StAR_CondDistr}
\end{equation}

(ii) Furthermore, for $t=1,2,\ldots$, the process $z_{t}$ has the representation
\begin{equation}
z_{t}=\varphi_{0}+\sum_{i=1}^{p}\varphi_{i}\thinspace z_{t-i}+\sigma_{t}\thinspace\varepsilon_{t}\label{eq:StAR_Model}
\end{equation}
with conditional variance $\sigma_{t}^{2}=\sigma^{2}(\boldsymbol{z}_{t-1})$
(see (\ref{eq:z_cond_mean_var})), where the error terms $\varepsilon_{t}$
form a sequence of independent and identically distributed random variables with a marginal $t_{1}(0,1,\nu+p)$
distribution and with $\varepsilon_{t}$ independent of $\{z_{s},\ s<t\}$. 
\end{thm}

Results (i) and (ii) in Theorem 1 are comparable to properties (\ref{eq:LinN_Distr})
and (\ref{eq:LinN_Model}) in the Gaussian case. Part (i) shows that
both the stationary and conditional distributions of $z_{t}$
are $t$\textendash distributions, whereas part (ii) clarifies the
connection to standard AR($p$) models. In contrast to linear Gaussian
autoregressions, in this $t$\textendash distributed case $z_{t}$
is conditionally heteroskedastic and has an `AR($p$)\textendash ARCH($p$)'
 representation (here ARCH refers to autoregressive conditional heteroskedasticity).

\section{A mixture autoregressive model based on Student's $t$\textendash distribution }

\subsection{Mixture autoregressive models}

Let $y_{t}$ ($t=1,2,\ldots$)
be the real-valued time series of interest, and let $\mathcal{F}_{t-1}$
denote the $\sigma$\textendash algebra generated by $\{y_{t-j},\text{ }j>0\}$.
We consider mixture autoregressive models for which the conditional
density function of $y_{t}$ given its past, $f(\cdot\mid\mathcal{F}_{t-1})$,
is of the form
\begin{equation}
f(y_{t}\mid\mathcal{F}_{t-1})=\sum_{m=1}^{M}\alpha_{m,t}f_{m}(y_{t}\mid\mathcal{F}_{t-1}),\label{Mixture AR general}
\end{equation}
where the (positive) mixing weights $\alpha_{m,t}$ are $\mathcal{F}_{t-1}$\textendash measurable
and satisfy $\sum_{m=1}^{M}\alpha_{m,t}=1$ (for all $t$), and the
$f_{m}(\cdot\mid\mathcal{F}_{t-1})$, $m=1,\ldots,M$, describe the
conditional densities of $M$ autoregressive component models. Different
mixture models are obtained with different specifications of the mixing
weights $\alpha_{m,t}$ and the conditional densities $f_{m}(\cdot\mid\mathcal{F}_{t-1})$.

Starting with the specification of the conditional densities $f_{m}(\cdot\mid\mathcal{F}_{t-1})$,
a common choice has been to assume the component models to be linear
Gaussian autoregressions. For the $m$th component model ($m=1,\ldots,M$),
denote the parameters of a $p$th order linear autoregression with
$\varphi_{m,0}\in\mathbb{R}$, $\boldsymbol{\varphi}_{m}=(\varphi_{m,1},\ldots,\varphi_{m,p})\in\mathbb{S}^{p}$,
and $\sigma_{m}>0$. Also set $\boldsymbol{y}_{t-1}=(y_{t-1},\ldots,y_{t-p})$.
In the Gaussian case, the conditional densities in (\ref{Mixture AR general})
take the form ($m=1,\ldots,M$) 
\[
f_{m}(y_{t}\mid\mathcal{F}_{t-1})=\frac{1}{\sigma_{m}}\phi\Bigl(\frac{y_{t}-\mu_{m,t}}{\sigma_{m}}\Bigr),
\]
where $\phi(\cdot)$ signifies the density function of a standard
normal random variable, $\mu_{m,t}=\varphi_{m,0}+\boldsymbol{\varphi}_{m}'\boldsymbol{y}_{t-1}$
is the conditional mean function (of component $m$), and $\sigma_{m}^{2}>0$
is the conditional variance (of component $m$), often assumed
to be constant. Instead of a Gaussian density, \citet{wong2009student}
consider the case where $f_{m}(\cdot\mid\mathcal{F}_{t-1})$ is the
density of Student's $t$\textendash distribution with conditional
mean and variance as above, $\mu_{m,t}=\varphi_{m,0}+\boldsymbol{\varphi}_{m}'\boldsymbol{y}_{t-1}$
and a constant $\sigma_{m}^{2}$, respectively. 

In this paper, we also consider a mixture autoregressive model based
on Student's $t$\textendash distribution, but our formulation differs
from that used by \citet{wong2009student}. In Theorem 1 it was seen
that linear autoregressions based on Student's $t$\textendash distribution
naturally lead to the conditional distribution $t_{1}(\mu(\cdot),\sigma^{2}(\cdot),\nu+p)$
in (\ref{eq:StAR_CondDistr}). Motivated by this, we consider a mixture
autoregressive model in which the conditional densities $f_{m}(y_{t}\mid\mathcal{F}_{t-1})$
in (\ref{Mixture AR general}) are specified as
\begin{equation}
f_{m}(y_{t}\mid\mathcal{F}_{t-1})=t_{1}(y_{t};\mu_{m,t},\sigma_{m,t}^{2},\nu_{m}+p),\label{f_m t distr}
\end{equation}
where the expressions for $\mu_{m,t}=\mu_{m}(\boldsymbol{y}_{t-1})$
and $\sigma_{m,t}^{2}=\sigma_{m}^{2}(\boldsymbol{y}_{t-1})$ are as
in (\ref{eq:z_cond_mean_var}) except that
$\boldsymbol{z}$ is replaced with $\boldsymbol{y}_{t-1}$
and all the quantities therein are defined using the regime specific
parameters $\varphi_{m,0}$, $\boldsymbol{\varphi}_{m}$,
$\sigma_{m}$, and $\nu_{m}$ (whenever appropriate a subscript
$m$ is added to previously defined notation,
e.g.,
$\mu_{m}$
or $\mathbf{\Gamma}_{m,p}$). A key difference to the model of \citet{wong2009student}
is that the conditional variance of component $m$ is not constant
but a function of $\boldsymbol{y}_{t-1}$. An explicit expression
for the density in (\ref{f_m t distr}) can be obtained from the Appendix and is
\begin{equation}
f_{m}(y_{t}\mid\mathcal{F}_{t-1}) 
= C(\nu_{m})\sigma_{m,t}^{-1}
\Bigl(1+(\nu_{m}+p-2)^{-1}\Bigl(\frac{y_{t}-\mu_{m,t}}{\sigma_{m,t}}\Bigr)^{2}\Bigr)^{-\frac{1+\nu_{m}+p}{2}},
\label{f_m t distr explicit}
\end{equation}
where $C(\nu)=\frac{\Gamma\left((1+\nu+p)/2\right)}{\left(\pi(\nu+p-2)\right)^{1/2}\Gamma\left((\nu+p)/2\right)}$
(and $\Gamma(\cdot)$ signifies the gamma function).

Now consider the choice of the mixing weights $\alpha_{m,t}$ in (\ref{Mixture AR general}).
The most basic choice is to use constant mixing weights as in \citet{wong2000mixture} and \citet{wong2009student}. Several
different time-varying mixing weights have also been suggested, see,
e.g., \citet{wong2001logistic}, \citet{glasbey2001non},
\citet{lanne2003modeling}, \citet{dueker2007contemporaneous}, 
and \citet{kalliovirta2015gaussian,kalliovirta2016gaussian}. 

In this paper, we propose mixing weights that are similar to those
used by \citet{glasbey2001non} and \citet{kalliovirta2015gaussian}.
Specifically, we set 
\begin{equation}
\alpha_{m,t}=\frac{\alpha_{m}t_{p}(\boldsymbol{y}_{t-1};\mu_{m}\mathbf{1}_{p},\mathbf{\Gamma}_{m,p},\nu_{m})}{\sum_{n=1}^{M}\alpha_{n}t_{p}(\boldsymbol{y}_{t-1};\mu_{n}\mathbf{1}_{p},\mathbf{\Gamma}_{n,p},\nu_{n})},\label{Mixing weights}
\end{equation}
where the $\alpha_{m}\in(0,1)$, $m=1,\ldots,M$, are unknown parameters satisfying $\sum_{m=1}^{M}\alpha_{m}=1$. 
Note that the Student's $t$ density appearing in 
(\ref{Mixing weights})
corresponds to the stationary distribution in Theorem 1(i): If the $y_{t}$'s
were generated by a linear Student's $t$ autoregression described
in Section 2 (with a subscript $m$ added to all the notation therein),
the stationary distribution of $\boldsymbol{y}_{t-1}$ would be characterized
by $t_{p}(\boldsymbol{y}_{t-1};\mu_{m}\mathbf{1}_{p},\mathbf{\Gamma}_{m,p},\nu_{m})$.
Our definition of the mixing weights in (\ref{Mixing weights}) is
different from that used in \citet{glasbey2001non} and \citet{kalliovirta2015gaussian}
in that these authors employed the $n_{p}(\boldsymbol{y}_{t-1};\mu_{m}\mathbf{1}_{p},\mathbf{\Gamma}_{m,p})$
density (corresponding to the stationary distribution of a linear
Gaussian autoregression) instead of the Student's $t$ density $t_{p}(\boldsymbol{y}_{t-1};\mu_{m}\mathbf{1}_{p},\mathbf{\Gamma}_{m,p},\nu_{m})$
we use. 

\subsection{{\normalsize{}The Student's $t$ mixture autoregressive model }}

Equations (\ref{Mixture AR general}), (\ref{f_m t distr}), and (\ref{Mixing weights})
define a model we call the Student's $t$ mixture autoregressive,
or StMAR, model. When the autoregressive order $p$ or the number
of mixture components $M$ need to be emphasized we refer to an StMAR($p$,$M$)
model. We collect the unknown parameters of an StMAR model in the
vector $\boldsymbol{\theta}=(\boldsymbol{\vartheta}_{1},\ldots,\boldsymbol{\vartheta}_{M},\alpha_{1},\ldots,\alpha_{M-1})$
($(M(p+4)-1)\times1$), where $\boldsymbol{\vartheta}_{m}=(\varphi_{m,0},\mathbf{\boldsymbol{\varphi}}_{m},\sigma_{m}^{2},\nu_{m})$
(with $\boldsymbol{\mathbf{\varphi}}_{m}\in\mathbb{S}^{p}$, $\sigma_{m}^{2}>0$,
and $\nu_{m}>2$) contains the parameters of each component model
($m=1,\ldots,M$) and the $\alpha_{m}$'s are the parameters appearing
in the mixing weights (\ref{Mixing weights}); the parameter $\alpha_{M}$
is not included due to the restriction $\sum_{m=1}^{M}\alpha_{m}=1$.

The StMAR model can also be presented in an alternative (but equivalent)
form. To this end, let $P_{t-1}\left(\cdot\right)$ signify the
conditional probability of the indicated event given $\mathcal{F}_{t-1}$,
and let $\varepsilon_{m,t}$ be a sequence of independent and identically distributed random variables
with a $t_{1}(0,1,\nu_{m}+p)$ distribution such that $\varepsilon_{m,t}$
is independent of $\{y_{t-j},\ j>0\}$ ($m=1,\ldots,M$). Furthermore,
let $\boldsymbol{s}_{t}=(s_{1,t},\ldots,s_{M,t})$ be a sequence of
(unobserved) $M$\textendash dimensional random vectors such that,
conditional on $\mathcal{F}_{t-1}$, $\boldsymbol{s}_{t}$ and $\varepsilon_{m,t}$
are independent (for all $m$). The components of $\boldsymbol{s}_{t}$
are such that, for each $t$, exactly one of them takes the value
one and others are equal to zero, with conditional probabilities $P_{t-1}(s_{m,t}=1)=\alpha_{m,t}$,
$m=1,\ldots,M$. Now $y_{t}$ can be expressed as 
\begin{equation}
y_{t}=\sum_{m=1}^{M}s_{m,t}(\mu_{m,t}+\sigma_{m,t}\varepsilon_{m,t})=\sum_{m=1}^{M}s_{m,t}(\varphi_{m,0}+\boldsymbol{\varphi}_{m}^{\prime}
\boldsymbol{y}_{t-1}+\sigma_{m,t}\varepsilon_{m,t}),\label{Mixture tAR s_t}
\end{equation}
where $\sigma_{m,t}$ is as in (\ref{f_m t distr}). This formulation
suggests that the mixing weights $\alpha_{m,t}$ can be thought of
as (conditional) probabilities that determine which one of the $M$
autoregressive components of the mixture generates the observation
$y_{t}$.

It turns out that the StMAR model has some very attractive theoretical
properties; the carefully chosen conditional
densities in (\ref{f_m t distr}) and the mixing weights in (\ref{Mixing weights})
are crucial in obtaining these properties. 
The following theorem shows that there exists a choice of initial values $\boldsymbol{y}_{0}$
such that $\boldsymbol{y}_{t}$ is a stationary and ergodic Markov
chain. Importantly, an explicit expression for the stationary distribution
is also provided. 
\begin{thm}
Consider the StMAR\ process $y_{t}$ generated by (\ref{Mixture AR general}),
(\ref{f_m t distr}), and (\ref{Mixing weights}) (or (\ref{Mixture tAR s_t})
and (\ref{Mixing weights})) with the conditions $\boldsymbol{\varphi}_{m}\in\mathbb{S}^{p}$
and $\nu_{m}>2$ satisfied for all $m=1,\ldots,M$. Then $\boldsymbol{y}_{t}=(y_{t},\ldots,y_{t-p+1})$
($t=1,2,\ldots$) is a Markov chain on $\mathbb{R}^{p}$ with a stationary
distribution characterized by the density\vspace*{-1pt}
\[
f(\boldsymbol{y}; \boldsymbol{\theta})=\sum_{m=1}^{M}\alpha_{m}t_{p}(\boldsymbol{y};\mu_{m}\mathbf{1}_{p},\mathbf{\Gamma}_{m,p},\nu_{m}).\vspace*{-1pt}
\]
Moreover, $\boldsymbol{y}_{t}$ is ergodic. 
\end{thm}
The stationary distribution of $\boldsymbol{y}_{t}$ is a mixture
of $M$ $p$\textendash dimensional $t$\textendash distributions
with constant mixing weights $\alpha_{m}$%
. Hence, moments of the stationary distribution
of order smaller than $\min\left(\nu_{1},\ldots,\nu_{M}\right)$ exist
and are finite. As can be seen from the proof of Theorem 2 (in the Supplementary Material), the stationary distribution
of the
vector $(y_{t},\boldsymbol{y}_{t-1})$
is also a mixture of $M$ $t$\textendash distributions with density
of the same form, $\sum_{m=1}^{M}\alpha_{m}t_{p+1}(\mu_{m}\mathbf{1}_{p+1},\mathbf{\Gamma}_{m,p+1},\nu_{m})$.
Thus the mean, variance, and first $p$ autocovariances
of $y_{t}$ are
(here the connection
between 
$\gamma_{m,j}$ and $\mathbf{\Gamma}_{m,p+1}$ is as in (\ref{eq:NotationGamma}))
\[
\mu\overset{def}{=}E[y_{t}]=\sum_{m=1}^{M}\alpha_{m}\mu_{m},\quad \gamma_{j}\overset{def}{=}Cov[y_{t},y_{t-j}]=\sum_{m=1}^{M}\alpha_{m}\gamma_{m,j}+\sum_{m=1}^{M}\alpha_{m}(\mu_{m}-\mu)^{2},\ j=0,\ldots,p.
\]
Subvectors of $(y_{t},\boldsymbol{y}_{t-1})$ also have stationary
distributions that belong to the same family (but this does not hold for higher dimensional vectors such as $(y_{t+1},y_{t},\boldsymbol{y}_{t-1})$). 

The fact that an explicit expression for the stationary (marginal)
distribution of the StMAR model is available is not only convenient
but also quite exceptional among mixture autoregressive models or
other related nonlinear autoregressive models (such as threshold or
smooth transition models). Previously, similar results have been obtained
by \citet{glasbey2001non} and \citet{kalliovirta2015gaussian} in
the context of mixture autoregressive models that are of the same
form but based on the Gaussian distribution (for a few rather simple
first order examples involving other models, see \citet[Section 4.2]{tong2011threshold}).

From the definition of the model, the conditional mean and variance
of $y_{t}$ are obtained as
\begin{equation}
E[y_{t}\mid\mathcal{F}_{t-1}]=\sum_{m=1}^{M}\alpha_{m,t}\mu_{m,t},\quad Var[y_{t}\mid\mathcal{F}_{t-1}]=\sum_{m=1}^{M}\alpha_{m,t}\sigma_{m,t}^{2}+\sum_{m=1}^{M}\alpha_{m,t}\biggl(\mu_{m,t}-\sum_{n=1}^{M}\alpha_{n,t}\mu_{n,t}\biggr)^{2}.\hspace{-3pt}\label{eq:y_t cond mean var}
\end{equation}
Except for the different definition of the mixing weights, the conditional
mean is as in the Gaussian mixture autoregressive model of \citet{kalliovirta2015gaussian}.
This is due to the 
well-known
fact that in the multivariate $t$\textendash distribution
the conditional mean is of the same linear form as in the multivariate
Gaussian distribution. However, unlike in the Gaussian case, the conditional
variance of the multivariate $t$\textendash distribution is not constant.
Therefore, in (\ref{eq:y_t cond mean var}) we have the time-varying
variance component $\sigma_{m,t}^{2}$ which in the models
of \citet{kalliovirta2015gaussian} and \citet{wong2009student}
is constant (in the latter model the mixing weights are also constants).
In (\ref{eq:y_t cond mean var}) both the mixing weights $\alpha_{m,t}$
and the variance components $\sigma_{m,t}^{2}$ are functions of $\boldsymbol{y}_{t-1}$,
implying that the conditional variance exhibits nonlinear autoregressive conditional heteroskedasticity. 
Compared to the aforementioned previous models our model may therefore
be useful in applications where the data exhibits rather strong conditional
heteroskedasticity.%

\section{Estimation}

The parameters of an StMAR model can be estimated by the method of
maximum likelihood (details of the numerical optimization methods employed and of simulation experiments are available in the Supplementary Material).
As the stationary distribution of the StMAR
process is known it is even possible to make use of initial values
and construct the exact likelihood function and obtain exact maximum likelihood estimates.
Assuming the observed data $y_{-p+1},\ldots,y_{0},y_{1},\ldots,y_{T}$
and stationary initial values, the log-likelihood function takes the
form
\begin{equation}
L_{T}(\boldsymbol{\theta})=\log\bigg(\sum_{m=1}^{M}\alpha_{m}t_{p}(\boldsymbol{y}_{0};\mu_{m}\mathbf{1}_{p},\mathbf{\Gamma}_{m,p},\nu_{m})\bigg)+\sum_{t=1}^{T}l_{t}(\boldsymbol{\theta}),\label{log-likelihood}
\end{equation}
where
\begin{equation}
l_{t}(\boldsymbol{\theta})=\log\bigg(\sum_{m=1}^{M}\alpha_{m,t}t_{1}(y_{t};\mu_{m,t},\sigma_{m,t}^{2},\nu_{m}+p)\bigg).\label{log-likelihood2}
\end{equation}
An explicit expression for the density appearing in (\ref{log-likelihood2})
is given in (\ref{f_m t distr explicit}), and the notation for $\mu_{m,t}$
and $\sigma_{m,t}^{2}$ is explained after (\ref{f_m t distr}). 
Although not made explicit, $\alpha_{m,t}$, $\mu_{m,t}$, and $\sigma_{m,t}^{2}$, as well as the quantities $\mu_{m}$, $\boldsymbol{\gamma}_{m,p}$, and $\boldsymbol{\Gamma}_{m,p}$, depend on the parameter vector $\boldsymbol{\theta}$. 

In (\ref{log-likelihood}) it has been assumed that
the initial values
$\boldsymbol{y}_{0}$ are generated
by the stationary distribution. If this assumption seems inappropriate
one can condition on initial values and drop the first term on the
right hand side of (\ref{log-likelihood}). In what follows we assume
that estimation is based on this conditional log-likelihood, namely
$L_{T}^{(c)}(\boldsymbol{\theta})=T^{-1}\sum_{t=1}^{T}l_{t}(\boldsymbol{\theta})$
which we, for convenience, have also scaled with the sample size.
Maximizing $L_{T}^{(c)}(\boldsymbol{\theta})$ with respect to $\boldsymbol{\theta}$ yields the maximum likelihood estimator denoted
by $\hat{\boldsymbol{\theta}}_{T}$. 

The permissible parameter space of $\boldsymbol{\theta}$, denoted
by $\boldsymbol{\Theta}$, needs to be constrained in various ways.
The stationarity conditions $\boldsymbol{\varphi}_{m}\in\mathbb{S}^{p}$,
the positivity of the variances $\sigma_{m}^{2}$, and the conditions
$\nu_{m}>2$ ensuring existence of second moments are all assumed
to hold (for $m=1,\ldots,M$). Throughout we assume that the number
of mixture components $M$ is known, and this also entails the requirement
that the parameters $\alpha_{m}$ ($m=1,\ldots,M$) 
are strictly positive (and strictly less than
unity whenever $M>1$). Further restrictions are required to ensure
identification. Denoting the true parameter value by
$\boldsymbol{\theta}_{0}$ and assuming stationary initial values,
the condition needed is that $l_{t}(\boldsymbol{\theta})=l_{t}(\boldsymbol{\theta}_{0})$
almost surely only if $\boldsymbol{\theta}=\boldsymbol{\theta}_{0}$. An additional
assumption needed for this is
\begin{equation}
\alpha_{1}>\cdots>\alpha_{M}>0\text{ \ \ and \ \ }\boldsymbol{\vartheta}_{i}=\boldsymbol{\vartheta}_{j}\text{ only if }1\leq i=j\leq M.\label{Ident Cond}
\end{equation}
From a practical point of view this assumption is not restrictive
because what it essentially requires is that the $M$ component models
cannot be `relabeled' and the same StMAR model obtained. We summarize
the restrictions imposed on the parameter space as follows.
\begin{assumption}
The true parameter value $\boldsymbol{\theta}_{0}$ is an interior
point of $\boldsymbol{\Theta}$, where $\boldsymbol{\Theta}$ is a
compact subset of $\{\boldsymbol{\theta}=(\boldsymbol{\vartheta}_{1},\ldots,\boldsymbol{\vartheta}_{M},\alpha_{1},\ldots,\alpha_{M-1})\in\mathbb{R}^{M(p+3)}\times(0,1)^{M-1}:\boldsymbol{\varphi}_{m}\in\mathbb{S}^{p},\,\sigma_{m}^{2}>0,\,\text{and }\nu_{m}>2\,\,\text{for all }m=1,\ldots,M,\,\text{and (\ref{Ident Cond}) holds}\}$.
\end{assumption}
Asymptotic properties of the maximum likelihood estimator can now be established under
conventional high-level conditions. Denote
$\mathcal{I}(\boldsymbol{\theta})
=E\bigl[\frac{\partial l_{t}(\boldsymbol{\theta})}{\partial\boldsymbol{\theta}}\frac{\partial l_{t}(\boldsymbol{\theta})}{\partial\boldsymbol{\theta}'}\bigr]$
and $\mathcal{J}(\boldsymbol{\theta})=E\bigl[\frac{\partial^{2}l_{t}(\boldsymbol{\theta})}{\partial\boldsymbol{\theta}\partial\boldsymbol{\theta}'}\bigr]$.
\begin{thm}
Suppose $y_{t}$ is generated by the stationary and ergodic StMAR process
of Theorem 2 and that Assumption 1 holds. Then 
$\hat{\boldsymbol{\theta}}_{T}$
is strongly consistent,
i.e.,
$\hat{\boldsymbol{\theta}}_{T}\to\boldsymbol{\theta}_{0}$
almost surely. Suppose further that (i) $T^{1/2}\frac{\partial}{\partial\boldsymbol{\theta}}L_{T}^{(c)}(\boldsymbol{\theta}_{0})\overset{d}{\to}N(0,\mathcal{I}(\boldsymbol{\theta}_{0}))$
with $\mathcal{I}(\boldsymbol{\theta}_{0})$ finite and positive definite,
(ii) $\mathcal{J}(\boldsymbol{\theta}_{0})=-\mathcal{I}(\boldsymbol{\theta}_{0})$,
and (iii) $E\bigl[\sup_{\boldsymbol{\theta}\in\boldsymbol{\Theta}_{0}}\bigl|\frac{\partial^{2}l_{t}(\boldsymbol{\theta})}{\partial\boldsymbol{\theta}\partial\boldsymbol{\theta}'}\bigr|\bigr]<\infty$
for some $\boldsymbol{\Theta}_{0}$, a compact convex set contained
in the interior of $\boldsymbol{\Theta}$ that has $\boldsymbol{\theta}_{0}$
as an interior point. Then $T^{1/2}(\hat{\boldsymbol{\theta}}_{T}-\boldsymbol{\theta}_{0})\overset{d}{\rightarrow}N\bigl(0,-\mathcal{J}(\boldsymbol{\theta}_{0})^{-1}\bigr)$.
\end{thm}
Of the conditions
in this theorem, (i) states that a central limit theorem holds for
the score vector (evaluated at $\boldsymbol{\theta}_{0}$) and that
the information matrix is positive definite, (ii) is the information
matrix equality, and (iii) ensures the uniform convergence of the
Hessian matrix (in some neighbourhood of $\boldsymbol{\theta}_{0}$).
These conditions are standard but their verification may be tedious.

Theorem 3 shows that the conventional limiting distribution applies
to the maximum likelihood estimator $\hat{\boldsymbol{\theta}}_{T}$ which implies
the applicability of standard likelihood-based tests. It is worth
noting, however, that here a correct specification of the number of
autoregressive components $M$ is required. In particular, if the
number of component models is chosen too large then some parameters
of the model are not identified and, consequently, the result of Theorem
3 and the validity of the related tests break down. This particularly
happens when one tests for the number of component models. Such tests
for mixture autoregressive models with Gaussian conditional densities (see (\ref{Mixture AR general}))
are developed by \citet{meitz2017testing}. 
The testing problem is highly nonstandard %
and extending their results to the present case is beyond the scope of this paper.

Instead of formal tests, in our empirical application we use 
information criteria to infer which model fits the
data best. Similar approaches have also been used by \citet{wong2009student}
and others. Note that once the number of regimes is (correctly)
chosen, standard likelihood-based inference can be used to choose
regime-wise autoregressive orders and to test other hypotheses of
interest.

\section{Empirical example}\label{EE}

\begin{figure}[pt]
\includegraphics[width=39pc, height=15pc]{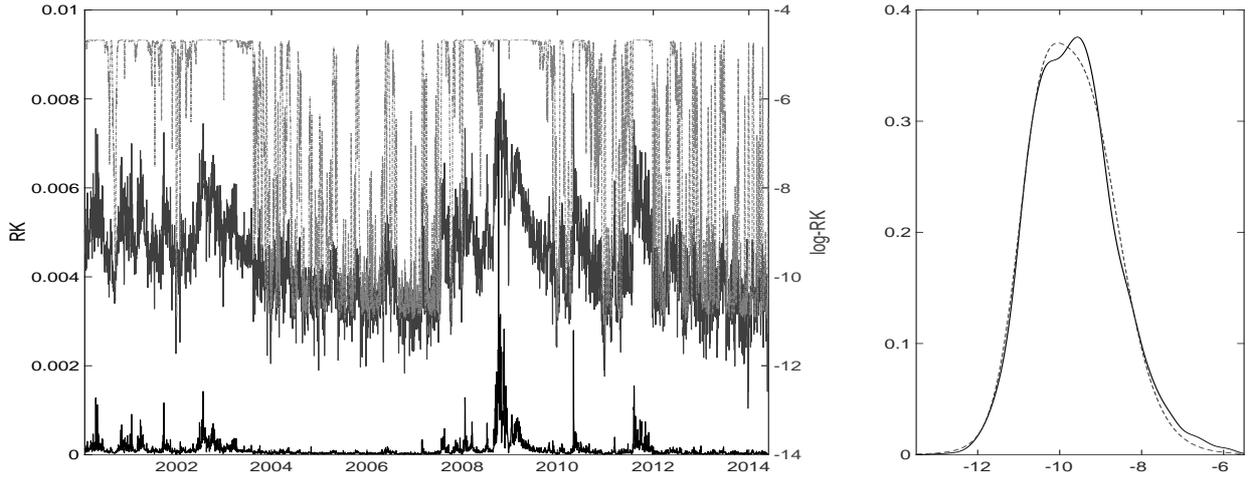}
\centering
\caption{Left panel: Daily $\textup{RK}_t$ (lower solid) and $\log(\textup{RK}_t)$ (upper solid), and mixing weights based on the estimates of the StMAR($4$,$2$) model in Table \ref{T1} (dot-dash) for the $\log(\textup{RK}_t)$ series. The mixing weights $\hat{\alpha}_{1,t}$ are scaled from $(0, \, 1)$ to $(\min \log(\textup{RK}_t), \, \max \log(\textup{RK}_t))$. Right panel: A kernel density estimate of the $\log(\textup{RK}_t)$ observations (solid), and the mixture density (dashes) implied by the same StMAR model as in the left panel.
}\label{F1}
\end{figure}

Modeling and forecasting financial market volatility is key to manage risk. In this application we use the realized kernel of \cite{barndorff2008designing} as a proxy for latent volatility. We obtained daily realized kernel data over the period 3 January 2000 through 20 May 2016 for the S\&P 500 index from the Oxford-Man Institute's Realized Library v0.2 \citep{heber2009oxford}. Figure \ref{F1} shows the in-sample period (Jan 3, 2000--June 3, 2014; 3597 observations) for the S\&P 500 realized kernel data ($\textup{RK}_t$), which is nonnegative with a distribution exhibiting substantial skewness and excess kurtosis (sample skewness 14.3, sample kurtosis 380.8). We follow the related literature which frequently use logarithmic realized kernel ($\log(\textup{RK}_t)$), to avoid imposing additional parameter constraints, and to obtain a more symmetric distribution, often taken to be approximately Gaussian. The $\log(\textup{RK}_t)$ data, also shown in Figure \ref{F1}, has a sample skewness of 0.5 and kurtosis of 3.5. Visual inspection of the time series plots of the $\textup{RK}_t$ and $\log(\textup{RK}_t)$ data suggests that the two series exhibit changes at least in levels and potentially also in variability. A kernel estimate of the density function of the $\log(\textup{RK}_t)$ series also suggest the potential presence of multiple regimes.

Table \ref{T1} reports estimation results for three selected StMAR models (for further details, see the Supplementary Material). Following  \cite{wong2001logistic}, \cite{wong2009student}, and \cite{li2015hysteretic}, we use information criteria for model comparison. For the $\log(\textup{RK}_t)$ data in-sample period the Akaike information criterion (\textsc{aic}) favours the StMAR($4$,$3$) model, the Hannan-Quinn information criterion (\textsc{hqc}) the StMAR($4$,$2$) model, and the Bayesian information criterion (\textsc{bic}) the simpler StMAR($4$,$1$) model. In view of the approximate standard errors in Table \ref{T1}, the estimation accuracy appears quite reasonable except for the degrees of freedom parameters. Taking the sum of the autoregressive parameters as a measure of persistence, we find that the estimated persistence for the first regime of the StMAR($4$,$2$) is 0.909 and 0.489 for the second regime, suggesting that persistence is rather strong in the first
regime and moderate in the second
regime.

\begin{table}[tbp]
\def~{\hphantom{0}}
\caption{%
Parameter
estimates for three selected StMAR models and the $\log(\textup{RK}_t)$ data over the period 3 January 2000 -- 3 June 2014. Numbers in parentheses are standard errors based on a numerical Hessian.}
\begin{center} 
\begin{tabular}{lcrrcrrcrr}
 \\
& & \multicolumn{2}{c}{StMAR($4,1$)} & & \multicolumn{2}{c}{StMAR($4,2$)} & & \multicolumn{2}{c}{StMAR($4,3$)} \\ [1.75 ex]
$\varphi_{1,0}$ & & $-0.746$ & ($0.089$)  & & $-0.851$ & ($0.112$) & & $-3.667$ & ($0.727$) \\ 
$\varphi_{1,1}$ & & $0.428$ & ($0.017$)  & & $0.432$ & ($0.024$) & & $0.331$ & ($0.035$) \\ 
$\varphi_{1,2}$ & & $0.224$ & ($0.019$)  & & $0.221$ & ($0.025$) & & $0.169$ & ($0.034$) \\ 
$\varphi_{1,3}$ & & $0.121$ & ($0.019$)  & & $0.122$ & ($0.025$) & & $0.055$ & ($0.033$) \\ 
$\varphi_{1,4}$ & & $0.150$ & ($0.017$)  & & $0.134$ & ($0.024$) & & $0.093$ & ($0.033$) \\ 
$\sigma_1^2$ & &  $0.298$ & ($0.011$)  & & $0.285$ & ($0.015$) & & $0.293$ & ($0.016$) \\ 
$\nu_1$ & & $11.999$ & ($1.109$)  & & $10.510$ & ($1.426$) & & $18.328$ & ($1.814$) \\ [5pt]
$\varphi_{2,0}$  & & & & & $-5.381$ & ($1.007$) & & $-1.013$ & ($0.341$) \\  
$\varphi_{2,1}$ & & & & & $0.289$ & ($0.046$) & & $0.509$ & ($0.038$) \\ 
$\varphi_{2,2}$ & & & & & $0.129$ & ($0.048$) & & $0.179$ & ($0.042$) \\ 
$\varphi_{2,3}$ & & & & & $0.023$ & ($0.046$) & & $0.043$ & ($0.045$) \\ 
$\varphi_{2,4}$ & & & & & $0.047$ & ($0.052$) & & $0.153$ & ($0.036$) \\ 
$\sigma_2^2$ & & & & & $0.287$ & ($0.022$) & & $0.327$ & ($0.024$) \\ 
$\nu_2$ & & & & & $29.031$ & ($1.595$) & & $12.977$ & ($2.200$) \\ [5pt]
$\varphi_{3,0}$ & & & & & & & & $-3.639$ & ($1.243$) \\  
$\varphi_{3,1}$ & & & & & & & & $0.208$ & ($0.072$) \\ 
$\varphi_{3,2}$ & & & & & & & & $0.198$ & ($0.082$) \\ 
$\varphi_{3,3}$ & & & & & & & & $0.219$ & ($0.067$) \\ 
$\varphi_{3,4}$ & & & & & & & & $-0.010$ & ($0.079$) \\ 
$\sigma_3^2$ & & & & & & & & $0.167$ & ($0.025$) \\ 
$\nu_3$ & & & & & & & & $22.008$ & ($2.697$) \\ [5pt]
$\alpha_1$ & & & & & $0.724$ & ($0.064$) & & $0.459$ & ($0.088$) \\  
$\alpha_2$ & & & & & & & & $0.342$ & ($0.099$) \\  [5pt]
$T\hspace{1pt}L_T^{(c)}(\hat{\boldsymbol{\theta}}_T)$ & & \multicolumn{2}{r}{$-2854.153$} & & \multicolumn{2}{r}{$-2832.665$} & & \multicolumn{2}{r}{$-2820.077$} \\ [5pt]
\textsc{aic} & & \multicolumn{2}{r}{$5722.306$} & & \multicolumn{2}{r}{$5695.330$} & & \multicolumn{2}{r}{$5686.154$} \\ 
\textsc{hqc} & & \multicolumn{2}{r}{$5737.741$} & & \multicolumn{2}{r}{$5728.406$} & & \multicolumn{2}{r}{$5736.870$} \\ 
\textsc{bic} & & \multicolumn{2}{r}{$5765.613$} & & \multicolumn{2}{r}{$5788.131$} & & \multicolumn{2}{r}{$5828.449$} 
\end{tabular}
\end{center} 
\label{T1}
\end{table}

\begin{table}
\def~{\hphantom{0}}
\caption{The percentage shares of cumulative realized kernel observations that belong to the 99\%, 95\% and 90\% one-sided upper prediction intervals based on the distribution of $500,000$ simulated conditional sample paths.}
\begin{center}
\begin{tabular}{lcrcrcrcrcrccr}
\\
 & & \multicolumn{5}{c}{\emph{Daily}} & & & \multicolumn{5}{c}{\emph{Weekly}} \\  [3pt] 
 & & \multicolumn{1}{c}{$99\%$} & & \multicolumn{1}{c}{$95\%$} & & \multicolumn{1}{c}{$90\%$} & & &  \multicolumn{1}{c}{$99\%$} & & \multicolumn{1}{c}{$95\%$} & & \multicolumn{1}{c}{$90\%$} \\ [5pt] 
AR($11$) & & $98.99$ & & $95.97$ & & $90.52$ & & & $96.54$ & & $91.26$ & & $86.18$ \\ 
HAR & & $98.59$ & & $94.76$ & & $90.52$ & & & $96.14$ & & $91.06$ & & $86.99$ \\ 
StMAR($4$,$1$) & & $98.99$ & & $95.97$ & & $92.14$ & & & $98.17$ & & $95.12$ & & $90.24$ \\ 
StMAR($4$,$2$) & &  $99.19$ & & $95.97$ & & $92.54$ & & & $97.97$ & & $94.92$ & & $90.65$ \\ 
StMAR($4$,$3$) & & $99.19$ & & $96.37$ & & $92.94$ & & & $98.37$ & & $94.72$ & & $90.65$ \\ [5pt]
 & & \multicolumn{5}{c}{\emph{Biweekly}} & & & \multicolumn{5}{c}{\emph{Monthly}} \\  [3pt] 
 & & \multicolumn{1}{c}{$99\%$} & & \multicolumn{1}{c}{$95\%$} & & \multicolumn{1}{c}{$90\%$} & & &  \multicolumn{1}{c}{$99\%$} & & \multicolumn{1}{c}{$95\%$} & & \multicolumn{1}{c}{$90\%$} \\ [5pt] 
AR($11$) & & $94.05$ & & $89.53$ & & $85.63$ & & & $94.11$ & & $88.63$ & & $85.47$ \\ 
HAR & & $93.63$ & & $88.71$ & & $84.80$ & & & $91.79$ & & $87.37$ & & $83.79$ \\ 
StMAR($4$,$1$) & & $97.33$ & & $93.22$ & & $90.76$ & & & $97.89$ & & $93.89$ & & $91.79$ \\ 
StMAR($4$,$2$) & & $97.33$ & & $93.22$ & & $90.76$ & & & $97.26$ & & $94.11$ & & $91.16$ \\ 
StMAR($4$,$3$) & & $97.54$ & & $93.22$ & & $90.97$ & & & $97.89$ & & $94.32$ & & $91.37$ 
\end{tabular}
\end{center}
\label{T3}
\end{table}

Numerous alternative models
for
volatility proxies have been proposed. We employ Corsi's (2009) \nocite{corsi2009simple} heterogeneous autoregressive (HAR) model as it is arguably the most popular reference model for forecasting proxies such as the realized kernel. We also consider a $p$th-order autoregression as the AR($p$) often performs well in volatility proxy forecasting. The StMAR models are estimated using maximum likelihood, and the reference AR and HAR models by ordinary least squares. We use a fixed scheme, where the parameters of our volatility models are estimated just once using data from Jan 3, 2000--June 3, 2014.
These estimates are then used to generate all forecasts. The remaining 496 observations of our sample are used to compare the forecasts from the alternative models. As discussed in \cite{kalliovirta2016gaussian}, computing multi-step-ahead forecasts for mixture models like the StMAR is rather complicated. For this reason we use computer driven forecasts to predict future volatility: For each out-of-sample date $T$, and for each alternative model, we simulate 500,000 sample paths. Each path is of length 22 (representing one trading month) and conditional on the information available at date $T$. In these simulations unknown parameters are replaced by their estimates. As the simulated paths are for $\log(\textup{RK}_t)$, and our object of interest is $\textup{RK}_t$, an exponential transformation is applied. 

We examine daily, weekly (5 day), biweekly (10 day), and monthly (22 day) volatility forecasts generated by the alternative models; for instance, the weekly volatility forecast at date $T$ is the forecast for $\textup{RK}_{T+1}+\cdots+\textup{RK}_{T+5}$ (the 5-day-ahead {\em cumulative} realized kernel).
Table \ref{T3} reports the percentage shares of (1, 5, 10, and 22-day) cumulative $\textup{RK}_t$ out-of-sample observations that belong to the 99\%, 95\%, and 90\% one-sided upper prediction intervals based on the distribution of the simulated sample paths; these upper prediction intervals for volatility are related to higher levels of risk in financial markets. Overall, it is seen that the empirical coverage rates of the StMAR based prediction intervals are closer to the nominal levels than the ones obtained with the reference models. By comparison, the accuracy of the prediction intervals obtained with the popular HAR model quickly degrade as the forecast period increases.
The StMAR model performs well also when two-sided prediction intervals and point forecast accuracy are considered (for details, see the Supplementary Material).

\section*{Acknowledgement}
The authors thank the Academy of Finland for financial support.

\section*{Supplementary material}
\label{SM}
The supplementary material includes proofs of Theorems 1--3, information on the numerical optimization methods employed for maximum likelihood estimation, simulation experiments, and further details of the empirical example.


\section*{Appendix}
\subsection*{Properties of the multivariate Student's $t$\textendash distribution}

The standard form of the density function of the multivariate
Student's $t$\textendash distribution with $\nu$ degrees of freedom
and dimension $d$ is 
(see, e.g., \citet[p. 1]{kotz2004multivariate})
\[
f\left(\boldsymbol{x}\right)
=
\frac{\Gamma\left((d+\nu)/2\right)}{\left(\pi\nu\right)^{d/2}\Gamma\left(\nu/2\right)}
\det\left(\mathbf{\Sigma}\right)^{-1/2}\left(1+\nu^{-1}(\boldsymbol{x}-\boldsymbol{\mu})^{\prime}\mathbf{\Sigma}^{-1}(\boldsymbol{x}-\boldsymbol{\mu})\right)^{-\frac{d+\nu}{2}},
\]
where $\Gamma\left(\cdot\right)$ is the gamma function and $\boldsymbol{\mu}\in\mathbb{R}^{d}$ and $\mathbf{\Sigma}$ ($d\times d$),
a symmetric positive definite matrix, are parameters. For a random
vector $\mathbf{X}$ possessing this density, the mean and covariance
are $E[\mathbf{X}]=\boldsymbol{\mu}$
and $Cov[\mathbf{X}]=\mathbf{\Gamma}=\frac{\nu}{\nu-2}\mathbf{\Sigma}$
(assuming $\nu>2$). The density can 
be expressed in terms of $\boldsymbol{\mu}$ and $\mathbf{\Gamma}$
as 
\[
f\left(\boldsymbol{x}\right)=
\frac{\Gamma\left((d+\nu)/2\right)}{\left(\pi(\nu-2)\right)^{d/2}\Gamma\left(\nu/2\right)}
\det\left(\mathbf{\Gamma}\right)^{-1/2}\left(1+(\nu-2)^{-1}(\boldsymbol{x}-\boldsymbol{\mu})^{\prime}\mathbf{\Gamma}^{-1}(\boldsymbol{x}-\boldsymbol{\mu})\right)^{-\frac{d+\nu}{2}}.
\]
This form of the density function, denoted by $t_{d}(\boldsymbol{x};\boldsymbol{\mu},\mathbf{\Gamma},\nu)$,
is used in this paper, and the notation $\mathbf{X}\sim t_{d}(\boldsymbol{\mu},\mathbf{\Gamma},\nu)$
is used for a random vector $\mathbf{X}$ possessing this density.
Condition $\nu>2$ and positive definiteness of $\mathbf{\Gamma}$
will be tacitly assumed.

For marginal and conditional distributions, partition $\mathbf{X}$
as $\mathbf{X}=(\mathbf{X}_{1},\mathbf{X}_{2})$ where the components
have dimensions $d_{1}$ and $d_{2}$ ($d_{1}+d_{2}=d$). Conformably
partition $\boldsymbol{\mu}$ and $\mathbf{\Gamma}$ as $\boldsymbol{\mu}=(\boldsymbol{\mu}_{1},\boldsymbol{\mu}_{2})$
and 
\[
\mathbf{\Gamma}=\left[\begin{array}{cc}
\mathbf{\Gamma}_{11} & \mathbf{\Gamma}_{12}\\
\mathbf{\Gamma}_{12}^{\prime} & \mathbf{\Gamma}_{22}
\end{array}\right].
\]
Then the marginal distributions of $\mathbf{X}_{1}$ and $\mathbf{X}_{2}$
are $t_{d_{1}}(\boldsymbol{\mu}_{1},\mathbf{\Gamma}_{11},\nu)$ and
$t_{d_{2}}(\boldsymbol{\mu}_{2},\mathbf{\Gamma}_{22},\nu)$, respectively.
The conditional
distribution of $\mathbf{X}_{1}$ given $\mathbf{X}_{2}$ is also
a $t$\textendash distribution, namely
(see \citet[Sec. 2]{ding2016conditional}) 
\[
\mathbf{X}_{1}\mid(\mathbf{X}_{2}=\boldsymbol{x}_{2})\sim t_{d_{1}}(\boldsymbol{\mu}_{1\mid2}(\boldsymbol{x}_{2}),\mathbf{\Gamma}_{1\mid2}(\boldsymbol{x}_{2}),\nu+d_{2}),
\]
where $\boldsymbol{\mu}_{1\mid2}(\boldsymbol{x}_{2})=\boldsymbol{\mu}_{1}+\mathbf{\Gamma}_{12}\mathbf{\Gamma}_{22}^{-1}(\boldsymbol{x}_{2}-\boldsymbol{\mu}_{2})$
and $\mathbf{\Gamma}_{1\mid2}(\boldsymbol{x}_{2})=\frac{\nu-2+(\boldsymbol{x}_{2}-\boldsymbol{\mu}_{2})^{\prime}\mathbf{\Gamma}_{22}^{-1}(\boldsymbol{x}_{2}-\boldsymbol{\mu}_{2})}{\nu-2+d_{2}}(\mathbf{\Gamma}_{11}-\mathbf{\Gamma}_{12}\mathbf{\Gamma}_{22}^{-1}\mathbf{\Gamma}_{12}^{\prime})$.
Furthermore, 
$
t_{d}(\boldsymbol{x};\boldsymbol{\mu},\mathbf{\Gamma},\nu)=t_{d_{1}}(\boldsymbol{x}_{1};\boldsymbol{\mu}_{1\mid2}(\boldsymbol{x}_{2}),\mathbf{\Gamma}_{1\mid2}(\boldsymbol{x}_{2}),\nu+d_{2})\ t_{d_{2}}(\boldsymbol{x}_{2};\boldsymbol{\mu}_{2},\mathbf{\Gamma}_{22},\nu)
$.

Now consider a special case:\ a ($p+1$)\textendash dimensional random vector $\mathbf{X}\sim t_{p+1}(\mu\boldsymbol{1}_{p+1},\mathbf{\Gamma}_{p+1},\nu)$,
where $\mu\in\mathbb{R}$ and $\mathbf{\Gamma}_{p+1}$ is a symmetric
positive definite Toeplitz matrix. Note that the mean vector $\mu\mathbf{1}_{p+1}$
and the covariance matrix $\mathbf{\Gamma}_{p+1}$ have structures
similar to those of the mean and covariance matrix of a ($p+1$)\textendash dimensional
realization of a second order stationary process. More specifically,
assume that $\mathbf{\Gamma}_{p+1}$ is the covariance matrix of a
second order stationary AR($p$) process.

Partition $\mathbf{X}$ as $\mathbf{X}=\left(X_{1},\mathbf{X}_{2}\right)=\left(\mathbf{X}_{1},X_{p+1}\right)$
with $X_{1}$ and $X_{p+1}$ real valued and $\mathbf{X}_{1}$ and
$\mathbf{X}_{2}$ both $p\times1$ vectors. The marginal distributions of $\mathbf{X}_{1}$ and $\mathbf{X}_{2}$
are $\mathbf{X}_{1} \sim t_{p}(\mu\mathbf{1}_{p},\mathbf{\Gamma}_{p},\nu)$ and $\mathbf{X}_{2}\sim t_{p}(\mu\mathbf{1}_{p},\mathbf{\Gamma}_{p},\nu)$,
where the (symmetric positive definite Toeplitz) matrix $\mathbf{\Gamma}_{p}=Cov\left[\mathbf{X}_{1}\right]=Cov\left[\mathbf{X}_{2}\right]$
is obtained from $\mathbf{\Gamma}_{p+1}$ by deleting the first row
and first column or, equivalently, the last row and last column (here
the specific structures of $\mu\boldsymbol{1}_{p+1}$ and $\mathbf{\Gamma}_{p+1}$
are used). The conditional distribution of $X_{1}$ given
$\mathbf{X}_{2}=\boldsymbol{x}_{2}$ is 
\[
X_{1}\mid(\mathbf{X}_{2}=\boldsymbol{x}_{2})\sim t_{1}(\mu(\boldsymbol{x}_{2}),\sigma^{2}(\boldsymbol{x}_{2}),\nu+p),
\]
where expressions for $\mu(\boldsymbol{x}_{2})$ and $\sigma^{2}(\boldsymbol{x}_{2})$
can be obtained from above as follows. Partition $\mathbf{\Gamma}_{p+1}$
as 
\[
\mathbf{\Gamma}_{p+1}=\left[\begin{array}{cc}
\gamma_{0} & \mathbf{\boldsymbol{\gamma}}_{p}^{\prime}\\
\boldsymbol{\mathbf{\gamma}}_{p} & \mathbf{\Gamma}_{p}
\end{array}\right],
\]
and denote $\boldsymbol{\varphi}=\boldsymbol{\Gamma}_{p}^{-1}\boldsymbol{\gamma}_{p}$
and $\sigma^{2}=\gamma_{0}-\boldsymbol{\gamma}_{p}'\mathbf{\Gamma}_{p}^{-1}\boldsymbol{\gamma}_{p}$
($\sigma^{2}>0$ as $\mathbf{\Gamma}_{p+1}$ is 
positive definite). From above, 
\begin{eqnarray*}
\mu(\boldsymbol{x}_{2}) & = & \boldsymbol{\mu}_{1\mid2}(\boldsymbol{x}_{2})=\mu+\boldsymbol{\gamma}_{p}'\boldsymbol{\Gamma}_{p}^{-1}(\boldsymbol{x}_{2}-\mu\mathbf{1}_{p})=\mu(1-\boldsymbol{\gamma}_{p}'\boldsymbol{\Gamma}_{p}^{-1}\mathbf{1}_{p})+\boldsymbol{\varphi}'\boldsymbol{x}_{2},
\\
\sigma^{2}(\boldsymbol{x}_{2}) & = & \mathbf{\Gamma}_{1\mid2}(\boldsymbol{x}_{2})=\frac{\nu-2+(\boldsymbol{x}_{2}-\mu\mathbf{1}_{p})^{\prime}\mathbf{\Gamma}_{p}^{-1}(\boldsymbol{x}_{2}-\mu\mathbf{1}_{p})}{\nu-2+p}\sigma^{2}.
\end{eqnarray*}

\pagebreak{}


\setcounter{page}{1}
\setcounter{equation}{0}
\setcounter{table}{0}
\setcounter{figure}{0}

\renewcommand{\theequation}{A\arabic{equation}}

\section*{Supplementary material for \protect \\
``A mixture autoregressive model based on Student's $t$--distribution''\\
by Meitz, Preve, and Saikkonen}

This Supplementary Material includes proofs of Theorems 1--3, information on the numerical optimization methods employed for maximum likelihood estimation, simulation experiments, and further details of the empirical example.

\setcounter{section}{0}

\section{Proofs}

\begin{proof}[\textbf{\emph{Proof of Theorem 1}}]
 Corresponding to $\varphi_{0}\in\mathbb{R}$, $\boldsymbol{\varphi}=(\varphi_{1},\ldots,\varphi_{p})\in\mathbb{S}^{p}$,
$\sigma>0$, and $\nu>2$, define the notation $\mathbf{\Gamma}_{p}$,
$\gamma_{0}$, $\boldsymbol{\gamma}_{p}$, $\mu$, and $\mathbf{\Gamma}_{p+1}$
as in
(4)%
, and note that $\mathbf{\Gamma}_{p}$
and $\mathbf{\Gamma}_{p+1}$ are, by construction and due to assumption
$\boldsymbol{\varphi}\in\mathbb{S}^{p}$, symmetric positive definite
Toeplitz matrices. To prove (i), we will construct a $p$\textendash dimensional
Markov process $\boldsymbol{z}_{t}=(z_{t},\ldots,z_{t-p+1})$ ($t=1,2,\ldots$)
with the desired properties. We need to specify an appropriate transition
probability measure and an initial distribution. For the former, assume
that the transition probability measure of $\boldsymbol{z}_{t}$ is
determined by the density function $t_{1}(z_{t};\mu(\boldsymbol{z}_{t-1}),\sigma^{2}(\boldsymbol{z}_{t-1}),\nu+p)$,
where $\mu(\boldsymbol{z}_{t-1})$ and $\sigma^{2}(\boldsymbol{z}_{t-1})$
are obtained from 
the last two displayed equations in the Appendix
by substituting $\boldsymbol{z}_{t-1}$ for $\boldsymbol{x}_{2}$.
This shows that $\boldsymbol{z}_{t}$ can be treated as a Markov chain
(see 
Meyn and Tweedie (2009, Ch.\ 3)%
). Concerning the initial value
$\boldsymbol{z}_{0}$, suppose it follows the $t$\textendash distribution
$\boldsymbol{z}_{0}\sim t_{p}(\mu\mathbf{1}_{p},\mathbf{\Gamma}_{p},\nu)$.
Furthermore, if $\boldsymbol{z}_{t}^{+}=(z_{t},\boldsymbol{z}_{t-1})=(\boldsymbol{z}_{t},z_{t-p})$,
we find from
the Appendix
that the density
function of $\boldsymbol{z}_{1}^{+}$ is given by
\begin{equation}
t_{p+1}(\boldsymbol{z}_{1}^{+},\mu\boldsymbol{1}_{p+1},\mathbf{\Gamma}_{p+1},\nu)=t_{1}(z_{1};\mu(\boldsymbol{z}_{0}),\sigma^{2}(\boldsymbol{z}_{0}),\nu+p)\ t_{p}(\boldsymbol{z}_{0};\mu\mathbf{1}_{p},\mathbf{\Gamma}_{p},\nu).\label{AppB density decomposition}
\end{equation}
Thus, $\boldsymbol{z}_{1}^{+}\sim t_{p+1}(\mu\mathbf{1}_{p+1},\mathbf{\Gamma}_{p+1},\nu)$
and, as in the Appendix, it follows that the marginal distribution
of $\boldsymbol{z}_{1}$ is the same as that of $\boldsymbol{z}_{0}$,
that is, $\boldsymbol{z}_{1}\sim t_{p}(\mu\mathbf{1}_{p},\mathbf{\Gamma}_{p},\nu)$
(the specific structure of $\mathbf{\Gamma}_{p+1}$ is used here).
Hence, as $\boldsymbol{z}_{t}$ is a Markov chain, we can conclude
that it has a stationary distribution characterized by the density
function $t_{p}(\boldsymbol{z},\mu\mathbf{1}_{p},\mathbf{\Gamma}_{p},\nu)$
(see 
Meyn and Tweedie (2009, pp.\ 230--231)%
). This completes the proof
of (i).

To prove (ii), note that, due to the Markov property, $z_{t}\mid\mathcal{F}_{t-1}^{z}\sim t_{1}(\mu(\boldsymbol{z}_{t-1}),\sigma^{2}(\boldsymbol{z}_{t-1}),\nu+p)$
where $\mathcal{F}_{t-1}^{z}$ signifies the sigma-algebra generated
by $\{z_{s},s<t\}$. Thus we can write the conditional expectation
and conditional variance of $z_{t}$ given $\mathcal{F}_{t-1}^{z}$
as 
\begin{align*}
E[z_{t}\mid\mathcal{F}_{t-1}^{z}] & =E[z_{t}\mid\boldsymbol{z}_{t-1}]=\mu+\boldsymbol{\gamma}_{p}'\mathbf{\Gamma}_{p}^{-1}(\boldsymbol{z}_{t-1}-\mu\mathbf{1}_{p})=\varphi_{0}+\boldsymbol{\varphi}'\boldsymbol{z}_{t-1},\\
Var[z_{t}\mid\mathcal{F}_{t-1}^{z}] & =Var[z_{t}\mid\boldsymbol{z}_{t-1}]=\frac{\nu-2+(\boldsymbol{z}_{t-1}-\mu\mathbf{1}_{p})^{\prime}\mathbf{\Gamma}_{p}^{-1}(\boldsymbol{z}_{t-1}-\mu\mathbf{1}_{p})}{\nu-2+p}\sigma^{2}.
\end{align*}
Denote this conditional variance by $\sigma_{t}^{2}=\sigma^{2}(\boldsymbol{z}_{t-1})$
(and note that $\sigma_{t}^{2}>0$ a.s. due to the assumed conditions
$\sigma^{2}>0$, $\mathbf{\Gamma}_{p}>0$, and $\nu>2$). Now the
random variables $\varepsilon_{t}$ defined by
\[
\varepsilon_{t}\overset{def}{=}(z_{t}-\varphi_{0}-\boldsymbol{\varphi}'\boldsymbol{z}_{t-1})/\sigma_{t}
\]
follow, conditional on $\mathcal{F}_{t-1}^{z}$, the $t_{1}(0,1,\nu+p)$
distribution. Hence, we obtain the `AR($p$)\textendash ARCH($p$)'
representation 
(7)%
. Because the conditional distribution
$\varepsilon_{t}\mid\mathcal{F}_{t-1}^{z}\sim t_{1}(0,1,\nu+p)$ does
not depend on $\mathcal{F}_{t-1}^{z}$ (or, more specifically, on
the random variables $\{z_{s},\ s<t\}$), the same holds true also
unconditionally, $\varepsilon_{t}\sim t_{1}(0,1,\nu+p)$, implying
that the random variables $\varepsilon_{t}$ are independent of $\mathcal{F}_{t-1}^{z}$
(or of $\{z_{s},\ s<t\}$). Moreover, from the definition of the $\varepsilon_{t}$'s
it follows that $\{\varepsilon_{s},\ s<t\}$ is a function of $\{z_{s},\ s<t\}$,
and hence $\varepsilon_{t}$ is also independent of $\{\varepsilon_{s},\ s<t\}$.
Consequently, the random variables $\varepsilon_{t}$ are IID $t_{1}(0,1,\nu+p)$,
completing the proof of (ii).
\end{proof}

\smallskip{}

\begin{proof}[\textbf{\emph{Proof of Theorem 2}}]
 First note that $\boldsymbol{y}_{t}$ is a Markov chain on $\mathbb{R}^{p}$.
Now, let $\boldsymbol{y}_{0}=\left(y_{0},\ldots,y_{-p+1}\right)$
be a random vector whose distribution has the density $f(\boldsymbol{y}_{0};\boldsymbol{\theta})=\sum_{m=1}^{M}\alpha_{m}t_{p}(\boldsymbol{y}_{0};\mu_{m}\mathbf{1}_{p},\mathbf{\Gamma}_{m,p},\nu_{m})$.
According to 
(8), (9), (11),
and (\ref{AppB density decomposition}), the conditional density of
$y_{1}$ given $\boldsymbol{y}_{0}$ is
\begin{align*}
f(y_{1}\mid\boldsymbol{y}_{0};\boldsymbol{\theta}) & =\sum_{m=1}^{M}\frac{\alpha_{m}t_{p}(\boldsymbol{y}_{0};\mu_{m}\mathbf{1}_{p},\mathbf{\Gamma}_{m,p},\nu_{m})}{\sum_{n=1}^{M}\alpha_{n}t_{p}(\boldsymbol{y}_{0};\mu_{n}\mathbf{1}_{p},\mathbf{\Gamma}_{n,p},\nu_{n})}t_{1}(y_{1};\mu(\boldsymbol{y}_{0}),\sigma^{2}(\boldsymbol{y}_{0}),\nu_{m}+p)\\
 & =\sum_{m=1}^{M}\frac{\alpha_{m}}{\sum_{n=1}^{M}\alpha_{n}t_{p}(\boldsymbol{y}_{0};\mu_{n}\mathbf{1}_{p},\mathbf{\Gamma}_{n,p},\nu_{n})}t_{p+1}((y_{1},\boldsymbol{y}_{0});\mu_{m}\mathbf{1}_{p+1},\mathbf{\Gamma}_{m,p+1},\nu_{m}).
\end{align*}
It follows that the density of $(y_{1},\boldsymbol{y}_{0})$ is $f((y_{1},\boldsymbol{y}_{0});\boldsymbol{\theta})=\sum_{m=1}^{M}\alpha_{m}t_{p+1}((y_{1},\boldsymbol{y}_{0});\mu_{m}\mathbf{1}_{p+1},\mathbf{\Gamma}_{m,p+1},\nu_{m})$.
Integrating $y_{-p+1}$ out (and using the properties of marginal
distributions of a multivariate $t$\textendash distribution in the Appendix) shows that the density of $\boldsymbol{y}_{1}$ is $f(\boldsymbol{y}_{1};\boldsymbol{\theta})=\sum_{m=1}^{M}\alpha_{m}t_{p}(\boldsymbol{y}_{1};\mu_{m}\mathbf{1}_{p},\mathbf{\Gamma}_{m,p},\nu_{m})$.
Therefore, $\boldsymbol{y}_{0}$ and $\boldsymbol{y}_{1}$ are identically
distributed. As $\{\boldsymbol{y}_{t}\}_{t=1}^{\infty}$ is a (time
homogeneous) Markov chain, it follows that $\{\boldsymbol{y}_{t}\}_{t=1}^{\infty}$
has a stationary distribution $\pi_{\boldsymbol{y}}\left(\cdot\right)$,
say, characterized by the density $f(\cdot;\boldsymbol{\theta})=\sum_{m=1}^{M}\alpha_{m}t_{p}(\cdot;\mu_{m}\mathbf{1}_{p},\mathbf{\Gamma}_{m,p},\nu_{m})$
(cf.\ Meyn and Tweedie (2009, pp.\ 230--231)).

For ergodicity, let $P_{\boldsymbol{y}}^{p}(\boldsymbol{y},\cdot)=\Pr(\boldsymbol{y}_{p}\mid\boldsymbol{y}_{0}=\boldsymbol{y})$
signify the $p$\textendash step transition probability measure of
$\boldsymbol{y}_{t}$. It is straightforward to check that $P_{\boldsymbol{y}}^{p}(\boldsymbol{y},\cdot)$
has a density given by
\[
f(\boldsymbol{y}_{p}\mid\boldsymbol{y}_{0};\boldsymbol{\theta})=\prod_{t=1}^{p}f(y_{t}\mid\boldsymbol{y}_{t-1};\boldsymbol{\theta})=\prod_{t=1}^{p}\sum_{m=1}^{M}\mathsf{\alpha}_{m,t}t_{1}(y_{t};\mu(\boldsymbol{y}_{t-1}),\sigma^{2}(\boldsymbol{y}_{t-1}),\nu_{m}+p).
\]
The last expression makes clear that $f(\boldsymbol{y}_{p}\mid\boldsymbol{y}_{0};\boldsymbol{\theta})>0$
for all $\boldsymbol{y}_{p}\in\mathbb{R}^{p}$ and all $\boldsymbol{y}_{0}\in\mathbb{R}^{p}$.
Now, one can complete the proof that $\boldsymbol{y}_{t}$ is ergodic
in the sense of 
Meyn and Tweedie (2009, Ch.\ 13)
by using arguments
identical to those used in the proof of Theorem 1 in 
Kalliovirta et al.\ (2015).
\end{proof}

\smallskip{}

\begin{proof}[\textbf{\emph{Proof of Theorem 3}}]
 First note that Assumption 1 together with the continuity of $L_{T}^{(c)}(\boldsymbol{\theta})$
ensures the existence of a measurable maximizer $\hat{\boldsymbol{\theta}}_{T}$.
For strong consistency, it suffices to show that a certain uniform
convergence condition and a certain identification condition hold.
Specifically, the former required condition is that the conditional
log-likelihood function obeys a uniform strong law of large numbers,
that is, $\sup_{\boldsymbol{\theta\in\boldsymbol{\Theta}}}|L_{T}^{(c)}(\boldsymbol{\theta})-E[L_{T}^{(c)}(\boldsymbol{\theta})]|\to0$
a.s.~as $T\to\infty$. As the $y_{t}$'s are stationary and ergodic
and $E[L_{T}^{(c)}(\boldsymbol{\theta})]=E[l_{t}(\boldsymbol{\theta})]$,
condition $E\left[\sup_{\boldsymbol{\theta}\in\boldsymbol{\Theta}}\left|l_{t}(\boldsymbol{\theta})\right|\right]<\infty$
ensures that the uniform law of large numbers in 
Ranga Rao (1962)
applies.

The validity of condition $E\left[\sup_{\boldsymbol{\theta}\in\boldsymbol{\Theta}}\left|l_{t}(\boldsymbol{\theta})\right|\right]<\infty$ can be established by deriving suitable lower and upper bounds
for $l_{t}(\boldsymbol{\theta})$. Recall from 
(10) and (15)
that 
\[
l_{t}(\boldsymbol{\theta})=\log\bigg(\sum_{m=1}^{M}\alpha_{m,t}t_{1}(y_{t};\mu_{m,t},\sigma_{m,t}^{2},\nu_{m}+p)\bigg),
\]
where
\[
t_{1}(y_{t};\mu_{m,t},\sigma_{m,t}^{2},\nu_{m}+p)=C(\nu_{m})\sigma_{m,t}^{-1}\left(1+(\nu_{m}+p-2)^{-1}\Bigl(\frac{y_{t}-\mu_{m,t}}{\sigma_{m,t}}\Bigr)^{2}\right)^{-\frac{1+\nu_{m}+p}{2}}
\]
and $C(\nu)=\frac{\Gamma\left((1+\nu+p)/2\right)}{\left(\pi(\nu+p-2)\right)^{1/2}\Gamma\left((\nu+p)/2\right)}$.
The following arguments hold for some choice of finite positive constants
$c_{1},\ldots,c_{10}$, and all staments are understood to hold `for
all $m=1,\ldots,M$' whenever appropriate. The assumed compactness
of the parameter space (Assumption 1) and the continuity of the gamma
function on the positive real axis imply that 
\begin{equation}
c_{1}\leq C(\nu_{m})\leq c_{2}.\label{eq:Thm3SuppApp1}
\end{equation}
Next, recall that $\sigma_{m,t}^{2}=\frac{\nu_{m}-2+(\boldsymbol{y}_{t-1}-\mu_{m}\mathbf{1}_{p})^{\prime}\mathbf{\Gamma}_{m,p}^{-1}(\boldsymbol{y}_{t-1}-\mu_{m}\mathbf{1}_{p})}{\nu_{m}-2+p}\sigma_{m}^{2}$,
where the matrix $\mathbf{\Gamma}_{m,p}$ is positive definite and
$\sigma_{m}^{2}>0$. Thus, by the compactness of the parameter space,
$\sigma_{m,t}^{2}\geq c_{3}$. On the other hand, as $\mathbf{\Gamma}_{m,p}$
is a continuous function of the autoregressive coefficients, the continuity
of eigenvalues implies that the smallest eigenvalue of $\mathbf{\Gamma}_{m,p}$,
$\lambda_{min}(\mathbf{\Gamma}_{m,p})$, is bounded away from zero
by a constant. This, together with elementary inequalities, yields
$(\boldsymbol{y}_{t-1}-\mu_{m}\mathbf{1}_{p})^{\prime}\mathbf{\Gamma}_{m,p}^{-1}(\boldsymbol{y}_{t-1}-\mu_{m}\mathbf{1}_{p})\leq\lambda_{min}^{-1}(\mathbf{\Gamma}_{m,p})\Vert\boldsymbol{y}_{t-1}-\mu_{m}\mathbf{1}_{p}\Vert^{2}\leq c_{4}(1+y_{t-1}^{2}+\cdots+y_{t-p}^{2})$.
Thus, by the compactness of the parameter space, we have $c_{3}\leq\sigma_{m,t}^{2}\leq c_{5}(1+y_{t-1}^{2}+\cdots+y_{t-p}^{2})$
so that also
\begin{equation}
c_{5}^{-1}(1+y_{t-1}^{2}+\cdots+y_{t-p}^{2})^{-1}\leq\sigma_{m,t}^{-2}\leq c_{3}^{-1}.\label{eq:Thm3SuppApp2}
\end{equation}
Therefore 
\[
1\leq1+(\nu_{m}+p-2)^{-1}\Bigl(\frac{y_{t}-\mu_{m,t}}{\sigma_{m,t}}\Bigr)^{2}\leq c_{6}(1+y_{t}^{2}+y_{t-1}^{2}+\cdots+y_{t-p}^{2}),
\]
which, together with the compactness of the parameter space, implies
that 
\begin{equation}
c_{7}(1+y_{t}^{2}+y_{t-1}^{2}+\cdots+y_{t-p}^{2})^{-c_{8}}\leq\left(1+(\nu_{m}+p-2)^{-1}\Bigl(\frac{y_{t}-\mu_{m,t}}{\sigma_{m,t}}\Bigr)^{2}\right)^{-\frac{1+\nu_{m}+p}{2}}\leq1.\label{eq:Thm3SuppApp3}
\end{equation}
Using (\ref{eq:Thm3SuppApp1})\textendash (\ref{eq:Thm3SuppApp3})
it now follows that 
\[
c_{9}(1+y_{t-1}^{2}+\cdots+y_{t-p}^{2})^{-1/2}(1+y_{t}^{2}+y_{t-1}^{2}+\cdots+y_{t-p}^{2})^{-c_{8}}\leq t_{1}(y_{t};\mu_{m,t},\sigma_{m,t}^{2},\nu_{m}+p)\leq c_{10}.
\]
Using this and the fact that $\sum_{m=1}^{M}\alpha_{m,t}(\boldsymbol{\theta})=1$
we can now bound $l_{t}(\boldsymbol{\theta})$ from above by a constant,
say $l_{t}(\boldsymbol{\theta})\leq\bar{C}<\infty$. Furthermore,
for some $\underline{C}<\infty$, 
\[
-\underline{C}(1+\log(1+y_{t}^{2}+y_{t-1}^{2}+\cdots+y_{t-p}^{2}))\leq l_{t}(\boldsymbol{\theta}).
\]
Hence, as the StMAR process has finite second moments, we can conclude
that $E\left[\sup_{\boldsymbol{\theta}\in\boldsymbol{\Theta}}\left|l_{t}(\boldsymbol{\theta})\right|\right]<\infty$.

As for the latter condition required for consistency, we need to establish
that $E[l_{t}(\boldsymbol{\theta})]\leq E[l_{t}(\boldsymbol{\theta}_{0})]$
and that $E[l_{t}(\boldsymbol{\theta})]=E[l_{t}(\boldsymbol{\theta}_{0})]$
implies $\boldsymbol{\theta}=\boldsymbol{\theta}_{0}$. For notational
clarity, let us make the dependence on parameter values explicit in
the expressions in
(5)
and write $\mu(\cdot,\boldsymbol{\vartheta})$ and $\sigma^{2}(\cdot,\boldsymbol{\vartheta})$,
and let $\alpha_{m}(\boldsymbol{y},\boldsymbol{\theta})$ stand for
$\alpha_{m,t}$ (see 
(11)%
) but with $\boldsymbol{y}_{t-1}$
therein replaced by $\boldsymbol{y}$ and with the dependence on the
parameter values made explicit ($m=1,\ldots,M$). Making use of the
fact that the density of $(y_{t},\boldsymbol{y}_{t-1})$ has the form
$f((y_{t},\boldsymbol{y}_{t-1});\boldsymbol{\theta})=\sum_{m=1}^{M}\alpha_{m}t_{p+1}((y_{t},\boldsymbol{y}_{t-1});\mu_{m}\mathbf{1}_{p+1},\mathbf{\Gamma}_{m,p+1},\nu_{m})$
(see proof of Theorem 2) and reasoning based on the Kullback-Leibler
divergence, we can now use arguments analogous to those in 
Kalliovirta et al.\ (2015, p.\ 265)
to conclude that $E[l_{t}(\boldsymbol{\theta})]\leq E[l_{t}(\boldsymbol{\theta}_{0})]$
with equality if and only if for almost all $(y,\boldsymbol{y})$,
\begin{equation}
\sum_{m=1}^{M}\alpha_{m}(\boldsymbol{y},\boldsymbol{\theta})t_{1}(y;\mu(\boldsymbol{y},\boldsymbol{\vartheta}_{m}),\sigma^{2}(\boldsymbol{y},\boldsymbol{\vartheta}_{m}),\nu_{m}+p)=\sum_{m=1}^{M}\alpha_{m}(\boldsymbol{y},\boldsymbol{\theta}_{0})t_{1}(y;\mu(\boldsymbol{y},\boldsymbol{\vartheta}_{m,0}),\sigma^{2}(\boldsymbol{y},\boldsymbol{\vartheta}_{m,0}),\nu_{m,0}+p).\label{eq:ConsProof1}
\end{equation}
For each fixed $\boldsymbol{y}$ at a time, the mixing weights, conditional
means, and conditional variances in (\ref{eq:ConsProof1}) are constants,
and we may apply the results on identification of finite mixtures
of Student's $t$\textendash distributions in 
Holzmann et al.\ (2006, Example 1)
(their parameterization of the $t$\textendash distribution is slightly
different than ours, but identification with their parameterization
implies identification in our parameterization). Consequently, for
each fixed $\boldsymbol{y}$ at a time, there exists a permutation
$\{\tau(1),\ldots,\tau(M)\}$ of $\{1,\ldots,M\}$ (where this permutation
may depend on $\boldsymbol{y}$) such that
\begin{align}
\alpha_{m}(\boldsymbol{y},\boldsymbol{\theta}) & =\alpha_{\tau(m)}(\boldsymbol{y},\boldsymbol{\theta}_{0}),\,\,\,\mu(\boldsymbol{y},\boldsymbol{\vartheta}_{m})=\mu(\boldsymbol{y},\boldsymbol{\vartheta}_{\tau(m),0}),\,\,\,\sigma^{2}(\boldsymbol{y},\boldsymbol{\vartheta}_{m})=\sigma^{2}(\boldsymbol{y},\boldsymbol{\vartheta}_{\tau(m),0}),\text{\,\,and}\nonumber \\
 & \qquad\nu_{m}=\nu_{\tau(m),0}\text{\,\,for almost all }y\,\,(m=1,\ldots,M).\label{eq:ConsProof2}
\end{align}
The number of possible permutations being finite ($M!$), this induces
a finite partition of $\mathbb{R}^{p}$ where the elements $\boldsymbol{y}$
of each partition correspond to the same permutation. At least one
of these partitions, say $A\subset\mathbb{R}^{p}$, must have positive
Lebesque measure. Thus, (\ref{eq:ConsProof2}) holds for all fixed
$\boldsymbol{y}\in A$ with some specific permutation $\{\tau(1),\ldots,\tau(M)\}$
of $\{1,\ldots,M\}$. The fact that $\mu(\boldsymbol{y},\boldsymbol{\vartheta}_{m})=\mu(\boldsymbol{y},\boldsymbol{\vartheta}_{\tau(m),0})$
for $m=1,\ldots,M$, almost all $y$, and all $\boldsymbol{y}\in A$,
can be used to deduce that $(\varphi_{m,0},\boldsymbol{\varphi}_{m})=(\varphi_{m,0,0},\boldsymbol{\varphi}_{\tau(m),0})$
for $m=1,\ldots,M$ (see 
(4), (5), 
and
Kalliovirta et al.\ (2015, pp.\ 265--266)).
Similarly, using
condition $\sigma^{2}(\boldsymbol{y},\boldsymbol{\vartheta}_{m})=\sigma^{2}(\boldsymbol{y},\boldsymbol{\vartheta}_{\tau(m),0})$
(and the knowledge that $(\varphi_{m,0},\boldsymbol{\varphi}_{m},\nu_{m})=(\varphi_{m,0,0},\boldsymbol{\varphi}_{\tau(m),0},\nu_{m,0})$),
it follows that $\sigma_{m}^{2}=\sigma_{\tau(m),0}^{2}$ so that $\boldsymbol{\vartheta}_{m}=\boldsymbol{\vartheta}_{\tau(m),0}$
($m=1,\ldots,M$). Now $\alpha_{m}=\alpha_{\tau(m),0}$ ($m=1,\ldots,M$)
follows as in 
Kalliovirta et al.\ (2015, p.\ 266)).
In light of
(16),
the preceding facts imply that $\boldsymbol{\theta}=\boldsymbol{\theta}_{0}$.
This completes the proof of consistency. 

Given conditions (i)\textendash (iii) of the theorem, asymptotic normality
of the ML estimator can now be established using standard arguments.
The required steps can be found, for instance, in 
Kalliovirta et al.\ (2016, proof of Theorem 3).
We omit the details for brevity.
\end{proof}


\pagebreak

\section{Estimation}

\subsection{Numerical optimization}\label{NumOpt}

Finding maximum likelihood estimates of the unknown parameters of an StMAR($p$,$M$) model amounts to maximizing $L_{T}^{(c)}(\boldsymbol{\theta})$, a function in $M(p+4)-1$ variables, under several constraints. 
Our experience with both actual and simulated data indicates that this can be challenging, in part due to multiple local maxima, 
and that advanced numerical optimization methods are needed. 
We use a hybrid numerical optimization scheme combining randomized search methods and classical gradient based methods to efficiently search for a global maximum that satisfies the constraints. 

We first employ a genetic algorithm using a variety of initial populations (collections of starting points; for discussions on the genetic algorithm, other popular algorithms, and their applications in econometrics, see Goffe et al., 1994, and Dorsey and Mayer, 1995). For each of the initial populations, the genetic algorithm is run for a small number of generations to reach the region near an optimum point relatively quickly. 
Corresponding to each initial population, the solution found by the genetic algorithm is then used as a starting point for \textsc{Matlab}'s optimization method \texttt{fmincon}, which is faster and more efficient for local search (for \texttt{fmincon} we further use a sequential quadratic programming method; see e.g.\ Nocedal and Wright, 2006). The final parameter estimate is the best solution found by \texttt{fmincon} for all the starting points considered. This hybrid optimization scheme combining multiple initial populations, the genetic algorithm, and \texttt{fmincon} allows us to efficiently search the parameter space and reduces the risk of ending up with a local, not global, maximum. We parallelize our code to consider multiple initial populations and starting points in parallel. This helps to speed up the optimization considerably. In view of the complexity of the estimation procedure, numerical gradients and Hessians are used for the optimization.

The StMAR code used in our S\&P 500 realized kernel example, further described in our StMAR MATLAB Toolbox Documentation, is available for download through the second authors webpage at \href{https://www.researchgate.net/profile/Daniel_Preve}{https://www.researchgate.net/profile/Daniel\_Preve}. \textsc{R} code by Savi Virolainen is available through the CRAN repository in the form of the `\href{https://cran.r-project.org/web/packages/uGMAR/index.html}{uGMAR}' package.

\subsection{Simulation experiments} 

We carried out several Monte Carlo studies to evaluate the performance of the numerical optimization scheme described above. The results of two of these studies are reported in Tables \ref{MCtable1} and \ref{MCtable2}. In these experiments, $500$ independent simulated sample paths were generated from an StMAR($1$,$2$), and also from an StMAR($4$,$2$), process; the sample sizes and parameter values used are displayed in Tables \ref{MCtable1} and \ref{MCtable2}. 

Overall, the performance of the numerical optimization scheme is quite satisfactory. As is commonly known, the degrees of freedom parameter of a Student's $t$--distribution is relatively difficult to estimate, especially if its true value is large. This is also the case for our StMAR model, and our simulation results indicate that the $\nu_m$ parameters can be relatively difficult to estimate even in moderate or large samples. Similar difficulties were reported by Wong et al.\ (2009) when estimating their (constant mixing weights) version of a Student $t$-mixture autoregressive model using the EM algorithm (see their Table 3).

\vfill

\begin{table}[p]
\centering \caption{Simulation results for a StMAR($1$,$2$) with various sample sizes $T$ and $500$ replications.\newline M, Md and SD denote the sample mean, median, and standard deviation, respectively.} 
\begin{center}
\hspace*{-25pt}
\resizebox{1.0\textwidth}{!}{%
\begin{tabular}{lcrc rrr c rrr c rrr c rrr} 
 & & & & \multicolumn{3}{c}{$T = 500$} & & \multicolumn{3}{c}{$T = 1000$} & &  \multicolumn{3}{c}{$T = 2500$} & &  \multicolumn{3}{c}{$T = 5000$} \\  [1.25ex] 
 
 & & Value & & \makebox[25pt][c]{M}&\makebox[25pt][c]{Md}&\makebox[25pt][c]{SD} && \makebox[25pt][c]{M}&\makebox[25pt][c]{Md}&\makebox[25pt][c]{SD} && \makebox[25pt][c]{M}&\makebox[25pt][c]{Md}&\makebox[25pt][c]{SD} && \makebox[25pt][c]{M}&\makebox[25pt][c]{Md}&\makebox[25pt][c]{SD} \\  [1.5ex] 
 
$\varphi_{1,0}$ & & $-1.50$ 	&& $-2.31$ & $-1.75$ & $2.25$ 	&& $-2.02$ & $-1.58$ & $1.55$ 	&& $-1.64$ & $-1.51$ & $0.65$ 	&& $-1.52$ &  $-1.51$ &  $0.22$ \\ [0.5ex] 
$\varphi_{1,1}$ & & $ 0.85$ 	&& $0.73$ & $0.83$ & $0.20$ 		&& $0.78$ & $0.84$ & $0.16$ 		&& $0.83$ & $0.85$  & $0.08$ 		&& $0.85$ &  $0.85$ & $0.03$ \\ [0.5ex]  
$\sigma_1^2$ 	& & $ 0.35$ 	&& $0.38$ & $0.32$ & $0.45$ 		&& $0.35$ & $0.33$ & $0.09$ 		&& $0.35$ & $0.34$ & $0.05$ 		&& $0.35$ &  $0.35$ &  $0.04$ \\ [0.5ex]   
$\nu_1$ 		& & $ 4.00$ 	&& $14.01$ & $5.29$ & $44.43$ 	&& $6.45$ & $4.59$ & $11.47$ 		&& $4.47$ & $4.16$ & $2.22$ 		&& $4.12$ &  $4.03$ &  $0.47$ \\ [1.75ex]
 
$\varphi_{2,0}$ & & $-5.50$ 	&& $-4.79$ &  $-5.33$  & $2.92$ 	&& $-5.10$  & $-5.43$  & $1.65$ 	&& $-5.41$  & $-5.48$  & $0.84$ 	&& $-5.50$ &  $-5.49$ &  $0.39$ \\ [0.5ex] 
$\varphi_{2,1}$ & & $0.35$ 	&& $0.44$ &  $0.37$  & $0.30$ 	&& $0.40$  & $0.36$  & $0.21$ 	&& $0.36$  & $0.35$  & $0.10$ 	&&  $0.35$ &   $0.35$ &  $0.05$ \\ [0.5ex]    
$\sigma_2^2$  	& & $0.30$ 	&& $0.96$ &  $0.30$  & $14.03$ 	&& $0.31$  & $0.30$  & $0.06$ 	&& $0.30$  & $0.30$  & $0.03$ 	&& $0.30$ & $0.30$ & $0.02$ \\ [0.5ex]   
$\nu_2$ 		& & $8.00$ 	&& $20.48$ &  $7.08$  & $53.76$ 	&& $15.15$  & $7.44$  & $36.00$ 	&& $9.19$  & $7.76$  & $6.64$ 	&& $8.30$ &  $7.80$ &  $2.12$ \\ [1.75ex]
 
$\alpha_1$ 	& & $0.60$ 	&& $0.61$ &  $0.59$  & $0.10$ 	&& $0.59$  & $0.59$  & $0.07$ 	&& $0.59$  & $0.59$  & $0.04$ 	&& $0.60$ &  $0.60$ &  $0.03$ \\ [0.01ex]
\end{tabular}}\label{MCtable1}
\end{center}
\end{table}

\begin{table}[p]
\centering \caption{Simulation results for a StMAR($4$,$2$) with various sample sizes $T$ and $500$ replications.\newline M, Md and SD denote the sample mean, median, and standard deviation, respectively.} 
\begin{center}
\hspace*{-25pt}
\resizebox{1.0\textwidth}{!}{%
\begin{tabular}{lcrc rrr c rrr c rrr c rrr} 
 & & & & \multicolumn{3}{c}{$T = 1000$} & & \multicolumn{3}{c}{$T = 2500$} & &  \multicolumn{3}{c}{$T = 5000$} & &  \multicolumn{3}{c}{$T = 10000$} \\  [1.25ex] 
 
 & & Value & & \makebox[25pt][c]{M}&\makebox[25pt][c]{Md}&\makebox[25pt][c]{SD} && \makebox[25pt][c]{M}&\makebox[25pt][c]{Md}&\makebox[25pt][c]{SD} && \makebox[25pt][c]{M}&\makebox[25pt][c]{Md}&\makebox[25pt][c]{SD} && \makebox[25pt][c]{M}&\makebox[25pt][c]{Md}&\makebox[25pt][c]{SD} \\  [1.5ex] 
 
$\varphi_{1,0}$  && $-1.00$ 	&& $-1.65$  & $-1.16$  & $1.03$ 	&& $-1.46$  & $-1.08$  & $0.83$ 	&& $-1.27$  & $-1.03$  & $0.67$ 	&& $-1.15$  & $-1.02$  & $0.51$ \\ [0.5ex] 
$\varphi_{1,1}$  && $0.35$ 		&& $0.34$  & $0.34$  & $0.06$ 	&& $0.34$  & $0.34$  & $0.06$ 	&& $0.34$  & $0.35$  & $0.03$ 	&& $0.35$  & $0.35$  & $0.02$ \\ [0.5ex] 
$\varphi_{1,2}$  && $0.20$ 		&& $0.16$  & $0.17$  & $0.07$ 	&& $0.18$  & $0.19$  & $0.05$ 	&& $0.19$  & $0.20$  & $0.04$ 	&& $0.19$  & $0.20$  & $0.03$ \\ [0.5ex] 
$\varphi_{1,3}$  && $0.15$ 		&& $0.12$  & $0.13$  & $0.07$ 	&& $0.13$  & $0.14$  & $0.06$ 	&& $0.14$  & $0.15$  & $0.04$ 	&& $0.14$  & $0.15$  & $0.03$ \\ [0.5ex] 
$\varphi_{1,4}$  && $0.10$ 		&& $0.08$  & $0.08$  & $0.06$ 	&& $0.09$  & $0.09$  & $0.05$ 	&& $0.09$  & $0.09$  & $0.03$ 	&& $0.10$  & $0.10$  & $0.02$ \\ [0.5ex] 
$\sigma_1^2$   && $0.25$ 		&& $1.84$  & $0.26$  & $17.82$ 	&& $0.46$  & $0.25$  & $3.77$ 	&& $0.26$  & $0.25$  & $0.04$ 	&& $0.25$  & $0.25$  & $0.02$ \\ [0.5ex]   
$\nu_1$ 	        && $9.00$ 		&& $9.06$  & $7.02$  & $16.71$ 	&& $8.14$  & $8.32$  & $3.64$ 	&& $8.50$  & $8.87$  & $2.62$ 	&& $8.70$  & $8.88$  & $1.91$ \\ [1.75ex] 

$\varphi_{2,0}$  && $-3.00$ 	&& $-2.95$  & $-2.92$  & $2.86$ 	&& $-2.68$  & $-2.97$  & $0.92$ 	&& $-2.80$  & $-3.00$  & $0.72$ 	&& $-2.88$  & $-2.98$  & $0.53$ \\ [0.5ex] 
$\varphi_{2,1}$  && $0.30$ 		&& $0.29$  & $0.30$  & $0.14$ 	&& $0.30$  & $0.30$  & $0.04$ 	&& $0.30$  & $0.30$  & $0.03$ 	&& $0.30$  & $0.30$  & $0.02$ \\ [0.5ex]  
$\varphi_{2,2}$  && $0.10$ 		&& $0.11$  & $0.12$  & $0.13$ 		&& $0.12$  & $0.11$  & $0.06$ 		&& $0.11$  & $0.10$  & $0.04$ 		&& $0.11$  & $0.10$  & $0.03$ \\ [0.5ex]
$\varphi_{2,3}$  && $0.05$ 		&& $0.06$  & $0.07$  & $0.13$ 	&& $0.07$  & $0.06$  & $0.06$ 	&& $0.06$  & $0.05$  & $0.04$ 	&&  $0.06$  & $0.05$  & $0.03$ \\ [0.5ex]
$\varphi_{2,4}$  && $0.05$ 		&& $0.04$  & $0.04$  & $0.14$ 	&& $0.05$  & $0.05$  & $0.05$ 	&& $0.05$  & $0.05$  & $0.03$ 	&& $0.05$  & $0.05$  & $0.02$ \\ [0.5ex]
$\sigma_2^2$    && $0.30$ 		&& $1.32$  & $0.23$  & $12.28$ 	&& $0.35$  & $0.26$  & $0.42$ 	&& $0.32$  & $0.28$  & $0.17$ 	&& $0.30$  & $0.29$  & $0.07$ \\ [0.5ex]   
$\nu_2$             && $3.00$ 		&& $23.26$  & $4.63$  & $69.40$ 	&& $4.96$  & $3.38$  & $3.68$ 	&& $3.98$  & $3.11$  & $2.47$ 		&& $3.48$  & $3.04$  & $1.66$ \\ [1.75ex]

$\alpha_1$  && $0.55$ 		&& $0.62$  & $0.59$  & $0.10$ 	&& $0.57$  & $0.56$  & $0.05$ 	&& $0.56$  & $0.56$  & $0.04$ 	&& $0.55$  & $0.55$  & $0.03$ \\ [0.01ex]

\end{tabular}}\label{MCtable2}
\end{center}
\end{table}


\pagebreak
\section{Empirical example}

\subsection{In-sample results}\label{ISresults}

We estimated 12 different StMAR models with $p = 1, 2, 3, 4$ and $M = 1, 2, 3$. Of these models, the \textsc{bic}, \textsc{hqc}, and \textsc{aic} information criteria chose the StMAR($4$,$1$), StMAR($4$,$2$), and StMAR($4$,$3$) models, respectively. Estimation results for these three models are shown in Table 1 of the main paper.  
Higher-order models were also tried but their forecasting performance was inferior to the models with $p=4$.

\subsection{Out-of-sample results}\label{OSresults}

\subsubsection{Two-sided prediction intervals}

Table 2 of the main paper reported the percentage shares of 1, 5, 10, and 22-day cumulative $\textup{RK}_t$ out-of-sample observations that belong to the 99\%, 95\%, and 90\% one-sided upper prediction intervals based on the distribution of the simulated sample paths. 
The corresponding numbers for two-sided prediction intervals (for nominal levels 99\%, 95\%, 90\%, 70\%, and 50\%) are presented in Table \ref{AT3}.
Overall, it is seen that the empirical coverage rates of the StMAR based prediction intervals are closer to the nominal levels than the ones obtained with the reference models. The StMAR($4$,$1$), and also the StMAR($4$,$2$), does particularly well. By comparison, the accuracy of the prediction intervals obtained with the HAR quickly degrade as the forecast period increases. 

Note that to generate prediction intervals for the reference AR and HAR models, we need to specify an error distribution in these models; we assume that the errors are Gaussian. The order of the AR model is chosen using \textsc{aic} and \textsc{bic}; both favour an AR($p$) model with $p = 11$.

\vfill

\begin{table}[h]
    \centering \caption{The percentage shares of cumulative realized kernel observations that belong to the 99\%, 95\%, 90\%, 70\% and 50\% two-sided prediction intervals based on the distribution of 500,000 simulated conditional sample paths.} 
\begin{center} 
\hspace*{-15pt}
\resizebox{1.05\textwidth}{!}{%
\begin{tabular}{lcrcrcrcrcrccrcrcrcrcr}
 & & \multicolumn{9}{c}{\emph{Daily}} & & & \multicolumn{9}{c}{\emph{Weekly}} \\  [1.25ex] 
 & & \multicolumn{1}{c}{$99\%$} & & \multicolumn{1}{c}{$95\%$} & & \multicolumn{1}{c}{$90\%$} & & \multicolumn{1}{c}{$70\%$}  & & \multicolumn{1}{c}{$50\%$} & & &  \multicolumn{1}{c}{$99\%$} & & \multicolumn{1}{c}{$95\%$} & & \multicolumn{1}{c}{$90\%$} & & \multicolumn{1}{c}{$70\%$} & & \multicolumn{1}{c}{$50\%$}  \\ [1.5ex]
AR($11$) & & $98.39$ & & $94.35$ & & $89.92$ & & $65.52$ & & $45.97$ & & & $95.12$ & & $88.01$ & & $79.47$ & & $61.18$ & & $42.48$  \\ [0.25ex]
HAR & & $98.59$ & & $95.16$ & & $89.72$ & & $66.13$ & & $44.96$ & & & $94.31$ & & $86.79$ & & $79.67$ & & $60.37$ & & $41.87$  \\ [0.25ex] 
StMAR($4$,$1$) & & $98.79$ & & $94.35$ & & $88.91$ & & $64.92$ & & $45.56$ & & & $96.34$ & & $91.26$ & & $84.76$ & & $64.63$ & & $45.53$  \\ [0.25ex]
StMAR($4$,$2$) & & $98.79$ & & $94.56$ & & $89.31$ & & $66.53$ & & $48.39$ & & & $96.34$ & & $89.63$ & & $83.13$ & & $63.82$ & & $45.53$  \\ [0.25ex]
StMAR($4$,$3$) & & $98.59$ & & $93.95$ & & $89.11$ & & $66.13$ & & $47.38$ & & & $96.75$ & & $89.02$ & & $81.91$ & & $60.98$ & & $45.12$  \\ [2.5ex]
 & & \multicolumn{9}{c}{\emph{Biweekly}} & & & \multicolumn{9}{c}{\emph{Monthly}} \\  [1.25ex] 
 & & \multicolumn{1}{c}{$99\%$} & & \multicolumn{1}{c}{$95\%$} & & \multicolumn{1}{c}{$90\%$} & & \multicolumn{1}{c}{$70\%$}  & & \multicolumn{1}{c}{$50\%$} & & &  \multicolumn{1}{c}{$99\%$} & & \multicolumn{1}{c}{$95\%$} & & \multicolumn{1}{c}{$90\%$} & & \multicolumn{1}{c}{$70\%$} & & \multicolumn{1}{c}{$50\%$} \\ [1.5ex]
AR($11$) & & $93.22$ & & $83.98$ & & $77.82$ & & $60.99$ & & $43.94$ & & & $93.47$ & & $84.00$ & & $78.11$ & & $60.00$ & & $41.26$  \\ [0.25ex]
HAR & & $92.81$ & & $83.78$ & & $77.41$ & & $58.32$ & & $41.07$ & & & $90.11$ & & $80.84$ & & $76.42$ & & $56.63$ & & $38.32$  \\ [0.25ex] 
StMAR($4$,$1$) & & $96.92$ & & $89.94$ & & $84.39$ & & $65.71$ & & $48.46$ & & & $99.16$ & & $90.53$ & & $86.74$ & & $66.11$ & & $45.05$  \\ [0.25ex]
StMAR($4$,$2$) & & $96.71$ & & $88.50$ & & $81.52$ & & $62.01$ & & $44.15$ & & & $97.47$ & & $86.53$ & & $82.32$ & & $65.26$ & & $43.16$  \\ [0.25ex]
StMAR($4$,$3$) & & $96.10$ & & $86.04$ & & $79.06$ & & $59.55$ & & $42.71$ & & & $95.79$ & & $84.00$ & & $80.21$ & & $63.16$ & & $41.89$ \\ [2.75ex] 
\end{tabular}}\label{AT3}
\end{center}
\end{table}

\pagebreak

\subsubsection{Volatility point forecast evaluation criteria}

Let $RM$ denote a (cumulative) realized measure (volatility proxy), such as the realized variance or realized kernel, and $\widehat{RM}$ a forecast of $RM$. 
Although realized measures generally are consistent estimators of the underlying latent volatility, in practice they are noisy proxies. Because of this, care needs to be taken when choosing a loss function to evaluate and compare volatility forecasts. 
Following the literature on volatility forecast comparison (Patton and Sheppard, 2009; Patton, 2011), we consider the two most widely used loss functions, namely squared loss (MSE)
\[
	L_{MSE}( RM, \widehat{RM} ) = (RM - \widehat{RM})^2
\]
and QLIKE (quasi-likelihood) loss 
\[
	L_{QLIKE}( RM, \widehat{RM} ) = \frac{RM}{\widehat{RM}} - \log \frac{RM}{\widehat{RM}} - 1.
\]
Moreover, as Patton and Sheppard (2009) recommend the use of QLIKE rather than MSE in volatility forecasting applications, we employ QLIKE loss as our primary loss function, and squared loss as our secondary loss function.

\subsubsection{Point forecasts}

\begin{figure}[h]
	\includegraphics[width=0.70\textwidth]{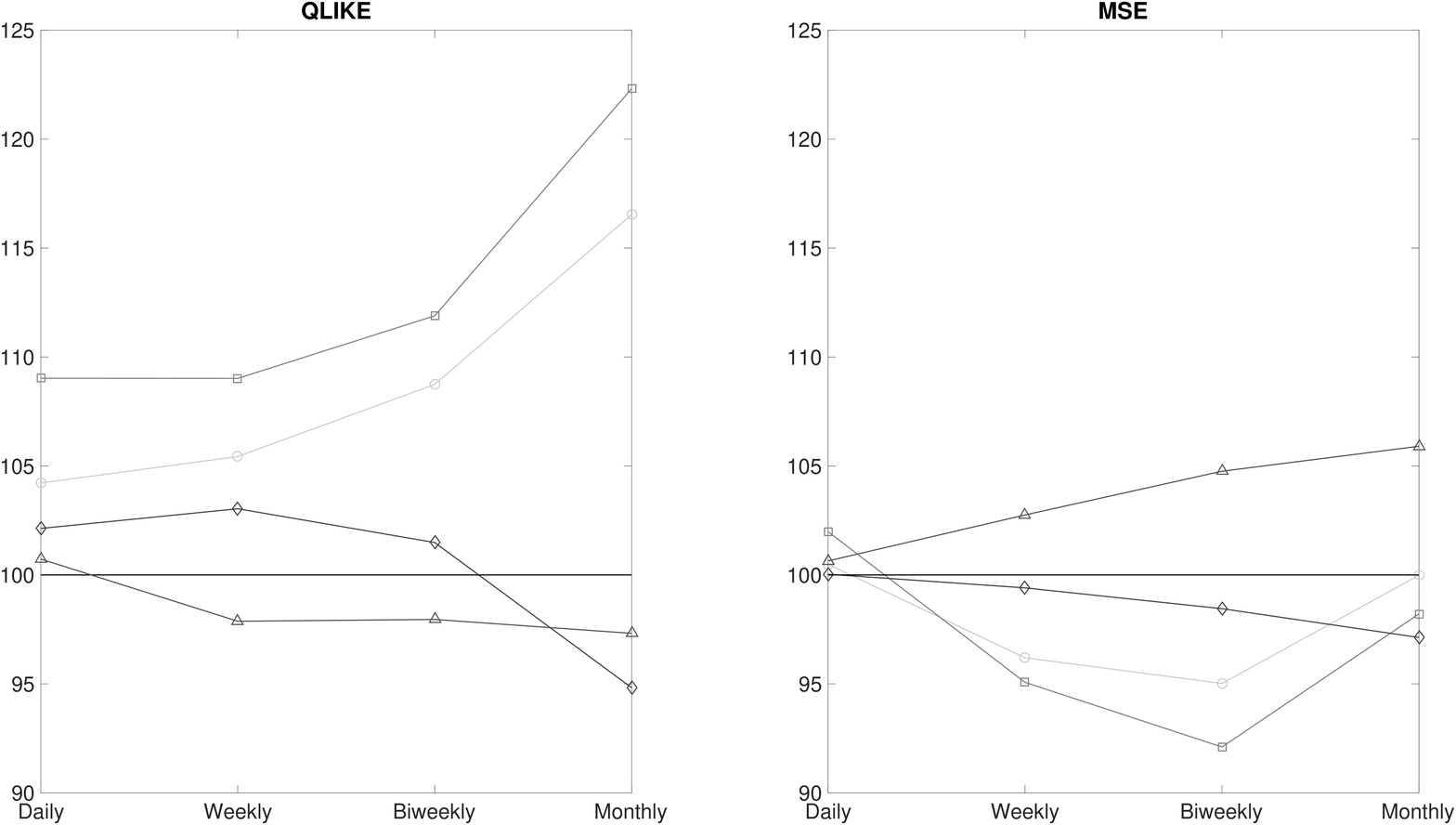}
\centering
\caption{Relative forecast accuracies for the S\&P 500 RK data in terms of QLIKE (left) and MSE (right). Results for the AR($11$) (circle), HAR (square), StMAR($4$,$1$) (diamond), StMAR($4$,$2$) (solid), and StMAR($4$,$3$) (triangle) models.}\label{F2}
\end{figure}

Results for 1, 5, 10, and 22-day cumulative $\textup{RK}_t$ forecasts based on the sample median are presented in Figure \ref{F2}. 
The left panel reports QLIKEs and the right panel MSEs. Forecast accuracy of the models is reported relative to the StMAR($4$,$2$) model: The horizontal line (at 100) represents the StMAR($4$,$2$) model, whereas the other lines represent the size of the forecast error measure made relative to this model (for instance, a value of 110 in the left panel is to be interpreted as a QLIKE 10\% larger than for the StMAR($4$,$2$) model). The overall performance of the StMAR($4$,$2$) model is quite reasonable. The model does particularly well in terms of our primary loss function, QLIKE. Overall, the StMAR($4$,$3$) performs somewhat more poorly in terms of MSE. Figure \ref{F2} also suggests that the more parsimonious StMAR($4$,$1$) model may be preferred to the StMAR($4$,$2$) model over longer (biweekly, monthly) forecast periods. The popular HAR model performs well under MSE, but considerably less so under QLIKE.

\pagebreak

\begin{spacing}{1.25}

\end{spacing}


\begin{thebibliography}{22}
\expandafter\ifx\csname natexlab\endcsname\relax\def\natexlab#1{#1}\fi

\bibitem[{Barndorff-Nielsen et~al.(2008)Barndorff-Nielsen, Hansen, Lunde \&
  Shephard}]{barndorff2008designing}
\textsc{Barndorff-Nielsen, O.~E.}, \textsc{Hansen, P.~R.}, \textsc{Lunde, A.}
  \& \textsc{Shephard, N.} (2008).
\newblock Designing realized kernels to measure the ex post variation of equity
  prices in the presence of noise.
\newblock \textit{Econometrica} \textbf{76}, 1481--1536.

\bibitem[{Corsi(2009)}]{corsi2009simple}
\textsc{Corsi, F.} (2009).
\newblock A simple approximate long-memory model of realized volatility.
\newblock \textit{J. Finan. Economet.} \textbf{7}, 174--196.

\bibitem[{Ding(2016)}]{ding2016conditional}
\textsc{Ding, P.} (2016).
\newblock On the conditional distribution of the multivariate $t$ distribution.
\newblock \textit{Am. Statistician} \textbf{70}, 293--295.

\bibitem[{Dueker et~al.(2007)Dueker, Sola \&
  Spagnolo}]{dueker2007contemporaneous}
\textsc{Dueker, M.~J.}, \textsc{Sola, M.} \& \textsc{Spagnolo, F.} (2007).
\newblock Contemporaneous threshold autoregressive models: estimation, testing
  and forecasting.
\newblock \textit{J. Economet.} \textbf{141}, 517--547.

\bibitem[{Fr{\"u}hwirth-Schnatter(2006)}]{fruhwirth2006finite}
\textsc{Fr{\"u}hwirth-Schnatter, S.} (2006).
\newblock \textit{Finite Mixture and Markov Switching Models}.
\newblock Springer.

\bibitem[{Glasbey(2001)}]{glasbey2001non}
\textsc{Glasbey, C.~A.} (2001).
\newblock Non-linear autoregressive time series with multivariate {G}aussian
  mixtures as marginal distributions.
\newblock \textit{J. R. Statist. Soc. C}
  \textbf{50}, 143--154.

\bibitem[{Heber et~al.(2009)Heber, Lunde, Shephard \&
  Sheppard}]{heber2009oxford}
\textsc{Heber, G.}, \textsc{Lunde, A.}, \textsc{Shephard, N.} \&
  \textsc{Sheppard, K.} (2009).
\newblock Oxford-man institute's realized library v0.2.
\newblock Oxford-Man Institute, University of Oxford.

\bibitem[{Heracleous \& Spanos(2006)}]{heracleous2006student}
\textsc{Heracleous, M.~S.} \& \textsc{Spanos, A.} (2006).
\newblock The {S}tudent's $t$ dynamic linear regression: re-examining
  volatility modeling.
\newblock In \textit{Econometric Analysis of Financial and Economic Time Series
  (Advaces in Econometrics, Vol 20 Part 1)}, D.~Terrell \& T.~B. Fomby, eds.
  Emerald Group Publishing Limited, pp. 289--319.

\bibitem[{Kalliovirta et~al.(2015)Kalliovirta, Meitz \&
  Saikkonen}]{kalliovirta2015gaussian}
\textsc{Kalliovirta, L.}, \textsc{Meitz, M.} \& \textsc{Saikkonen, P.} (2015).
\newblock A {G}aussian mixture autoregressive model for univariate time series.
\newblock \textit{J. Time Ser. Anal.} \textbf{36}, 247--266.

\bibitem[{Kalliovirta et~al.(2016)Kalliovirta, Meitz \&
  Saikkonen}]{kalliovirta2016gaussian}
\textsc{Kalliovirta, L.}, \textsc{Meitz, M.} \& \textsc{Saikkonen, P.} (2016).
\newblock Gaussian mixture vector autoregression.
\newblock \textit{J. Economet.} \textbf{192}, 485--498.

\bibitem[{Kotz \& Nadarajah(2004)}]{kotz2004multivariate}
\textsc{Kotz, S.} \& \textsc{Nadarajah, S.} (2004).
\newblock \textit{Multivariate $t$ distributions and their applications}.
\newblock Cambridge: Cambridge University Press.

\bibitem[{Lanne \& Saikkonen(2003)}]{lanne2003modeling}
\textsc{Lanne, M.} \& \textsc{Saikkonen, P.} (2003).
\newblock Modeling the {US} short-term interest rate by mixture autoregressive
  processes.
\newblock \textit{J. Finan. Economet.} \textbf{1}, 96--125.

\bibitem[{Le et~al.(1996)Le, Martin \& Raftery}]{le1996modeling}
\textsc{Le, N.~D.}, \textsc{Martin, R.~D.} \& \textsc{Raftery, A.~E.} (1996).
\newblock Modeling flat stretches, bursts, and outliers in time series using
  mixture transition distribution models.
\newblock \textit{J. Am. Statist. Assoc.} \textbf{91},
  1504--1515.

\bibitem[{Li et~al.(2015)Li, Guan, Li \& Yu}]{li2015hysteretic}
\textsc{Li, G.}, \textsc{Guan, B.}, \textsc{Li, W.~K.} \& \textsc{Yu, P.~L.}
  (2015).
\newblock Hysteretic autoregressive time series models.
\newblock \textit{Biometrika} \textbf{102}, 717--723.

\bibitem[{McLachlan \& Peel(2000)}]{mclachlan2000finite}
\textsc{McLachlan, G.} \& \textsc{Peel, D.} (2000).
\newblock \textit{Finite Mixture Models}.
\newblock Wiley.

\bibitem[{Meitz \& Saikkonen(2017)}]{meitz2017testing}
\textsc{Meitz, M.} \& \textsc{Saikkonen, P.} (2017).
\newblock Testing for observation-dependent regime switching in mixture
  autoregressive models.
\newblock {HECER} Discussion Paper No. 420, University of Helsinki,
  ar{X}iv:1711.03959.

\bibitem[{Spanos(1994)}]{spanos1994modeling}
\textsc{Spanos, A.} (1994).
\newblock On modeling heteroskedasticity: the {S}tudent's $t$ and elliptical
  linear regression models.
\newblock \textit{Economet. Theory} \textbf{10}, 286--315.

\bibitem[{Tong(2011)}]{tong2011threshold}
\textsc{Tong, H.} (2011).
\newblock Threshold models in time series analysis -- 30 years on.
\newblock \textit{Statistics and Its Interface} \textbf{4}, 107--118.

\bibitem[{Wong et~al.(2009)Wong, Chan \& Kam}]{wong2009student}
\textsc{Wong, C.~S.}, \textsc{Chan, W.~S.} \& \textsc{Kam, P.~L.} (2009).
\newblock A student $t$-mixture autoregressive model with applications to
  heavy-tailed financial data.
\newblock \textit{Biometrika} \textbf{96}, 751--760.

\bibitem[{Wong \& Li(2000)}]{wong2000mixture}
\textsc{Wong, C.~S.} \& \textsc{Li, W.~K.} (2000).
\newblock On a mixture autoregressive model.
\newblock \textit{J. R. Statist. Soc. B}
  \textbf{62}, 95--115.

\bibitem[{Wong \& Li(2001{\natexlab{a}})}]{wong2001logistic}
\textsc{Wong, C.~S.} \& \textsc{Li, W.~K.} (2001{\natexlab{a}}).
\newblock On a logistic mixture autoregressive model.
\newblock \textit{Biometrika} \textbf{88}, 833--846.

\bibitem[{Wong \& Li(2001{\natexlab{b}})}]{wong2001mixture}
\textsc{Wong, C.~S.} \& \textsc{Li, W.~K.} (2001{\natexlab{b}}).
\newblock On a mixture autoregressive conditional heteroscedastic model.
\newblock \textit{J. Am. Statist. Assoc.} \textbf{96},
  982--995.

\end{thebibliography}

\begin{thebibliography}{99}



\harvarditem[]{}{}{} Dorsey, R.E., and W.J.\ Mayer (1995) Genetic algorithms for estimation problems with multiple optima, nondifferentiability, and other irregular features. \textit{Journal of Business \& Economic Statistics} \textbf{13}(1): 53--66.

\harvarditem[]{}{}{} Goffe, W.L., G.D.\ Ferrier, and J.\ Rogers (1994) Global optimization of statistical functions with simulated annealing. \textit{Journal of Econometrics} \textbf{60}(1--2): 65--99.

\harvarditem[]{}{}{} Holzmann, H., A.\ Munk, and T.\  Gneiting (2006) Identifiability of finite mixtures of elliptical distributions. \textit{Scandinavian Journal of Statistics} \textbf{33}: 753--763.

\harvarditem[]{}{}{} Kalliovirta, L., M.\ Meitz, and P.\  Saikkonen (2015) A Gaussian mixture autoregressive model for univariate time series. \textit{Journal of Time Series Analysis} \textbf{36}: 247--266.

\harvarditem[]{}{}{} Kalliovirta, L., M.\ Meitz, and P.\  Saikkonen (2016) Gaussian mixture vector autoregression. \textit{Journal of Econometrics} \textbf{192}: 485--498.

\harvarditem[]{}{}{} Meyn, S., and R.L.\ Tweedie (2009) \textit{Markov Chains and Stochastic Stability (2nd ed.)}. Cambridge University Press, Cambridge.

\harvarditem[]{}{}{} Nocedal, J., and S.J.\ Wright (2006) \textit{Numerical Optimization (2nd ed.)}. Springer, New York.

\harvarditem[]{}{}{} Patton, A.J.\ (2011) Volatility forecast comparison using imperfect volatility proxies. \textit{Journal of Econometrics} \textbf{160}(1): 246--256. 

\harvarditem[]{}{}{} Patton, A.J., and K.\ Sheppard (2009) Evaluating volatility and correlation forecasts, in T.G.\ Andersen, R.A.\ Davis, J.P.\ Krei{\ss} and T.\ Mikosch (Eds.), \textit{Handbook of Financial Time Series}. Springer, Berlin Heidelberg.

\harvarditem[]{}{}{} Ranga Rao, R.\ (1962) Relations between weak and uniform convergence of measures with applications. \textit{Annals of Mathematical Statistics} \textbf{33}:659--680.

\harvarditem[]{}{}{} Wong, C.S., W.S.\ Chan, and P.L.\  Kam (2009) A Student $t$-mixture autoregressive model with applications to heavy-tailed financial data. \textit{Biometrika} \textbf{96}(3): 751--760.

\end{thebibliography}
\end{document}